\def\FOCS{0}

\ifnum\FOCS=1
\documentclass[times, 10pt,twocolumn]{article}
\usepackage{latex8}
\usepackage{times}
\usepackage{xspace}
\usepackage[final,notitlepage,nousetoc,nouselot,nouselof,nohylinks]{boaz}
\else
\documentclass[11pt]{article}
\usepackage{xspace}
\usepackage[final,notitlepage,nousetoc,nouselot,nouselof,nohylinks]{boaz}
\fi

\newcommand{\Mnote}[1]{{\authnote{Mohammad}{#1}}}
\newcommand{\abs}[1]{\lvert#1\rvert}

\newcommand{\tth}{^{\text{th}}} % it is like text phont but it is in the super index position
\newcommand{\SD}{\mathsf{SD}}

\newcommand{\Sign}{\mathsf{Sign}}
\newcommand{\Ver}{\mathsf{Ver}}
\newcommand{\Gen}{\mathsf{Gen}}
\newcommand{\Adv}{\mathsf{Adv}}

\newcommand{\cO}{\mathcal{O}}
\newcommand{\Hyb}{\mathbf{H}}

\usepackage{nicefrac}

\newcommand{\Supp}{\mathsf{Supp}}

\ifnum\FOCS=0
\let\Section\section
\let\SubSection\subsection

\else

\fi

\begin{document}

\title{Lower Bounds on Signatures From Symmetric Primitives}
\date{July 5, 2010}%\vspace{10pt} December 2012}

\ifnum\full=1
\author{Boaz Barak\thanks{Department of Computer Science, Princeton University.
Email: \texttt{boaz@cs.princeton.edu}} \and Mohammad
Mahmoody\thanks{Department of Computer Science,
Princeton University. Email:
\texttt{mohammad@cs.princeton.edu}}}

\else
\author{Boaz Barak  \hspace{4ex} Mohammad Mahmoody-Ghidary \\
Department of Computer Science \\
Princeton University \\
\texttt{\{boaz,mohammad\}@cs.princeton.edu}}

\fi

\begin{DOCheader}

\begin{abstract}
We show that every construction of one-time signature schemes
from a random oracle achieves black-box security at most
$2^{(1+o(1))q}$, where $q$ is the total number of oracle queries
asked by the key generation, signing, and verification
algorithms.  That is, any such scheme can be broken with
probability close to $1$ by a (computationally unbounded)
adversary making $2^{(1+o(1))q}$ queries to the oracle. This is
tight up to a constant factor in the number of queries, since a
simple modification of Lamport's one-time signatures (Lamport
'79) achieves $2^{(0.812-o(1))q}$ black-box security using $q$ queries to the oracle.
\Bnote{standard def of signature scheme is for any size message,
so there's no need to talk about $n$ in the abstract.}

Our result extends (with a loss of a constant factor in the
number of queries) also to the random permutation and
ideal-cipher oracles. Since the symmetric primitives (e.g. block
ciphers, hash functions, and message authentication codes) can
be constructed by a constant number of queries to the mentioned
oracles, as corollary we get lower bounds on the efficiency of
signature schemes from symmetric primitives when the
construction is black-box. This can be taken as evidence of an
inherent efficiency gap between signature schemes and symmetric
primitives.

\end{abstract}

\DOCkeywords{Cryptography, Digital Signatures, One-Time Signatures,
Black-Box Lower Bounds,  Random Oracle Model, Ideal Cipher Model}
\thispagestyle{empty}
\end{DOCheader}

\nocite{GennaroGeKaTr05}

\nnspace\Section{Introduction}

\emph{Digital signature schemes} allow authentication of
messages between parties without shared keys. Signature schemes
pose an interesting disconnect between the worlds of theoretical
and applied cryptography. From a theoretical point of view, it
is natural to divide cryptographic tools into those that can be
constructed using one-way functions and those that are not known
to have such constructions. Signature schemes, along with
private key encryption, message authentication codes,
pseudorandom generators and functions, belong to the former
camp. In contrast, the known constructions of \emph{public key
encryption} are based on \emph{structured} problems that are
conjectured to be hard (i.e., problems from number theory or the
theory of lattices).  {From} a practical point of view, it is
more natural to divide the tools according to the
\emph{efficiency} of their best known constructions. The
division is actually similar, since schemes based on structured
problems typically require both more complicated computations
and larger key size, as they often have non-trivial attacks
(e.g., because of the performance of the best known factoring
algorithms, to get $2^n$ security based on factorization one
needs to use $\Tilde{\Omega}(n^3)$ bit long integers).

Signature schemes are outlier to this rule: even though they can
be constructed using one-way functions, applied cryptographers
consider them as relatively inefficient since practical
constructions are based on structured hard problems, and thus
are significantly less efficient than private key encryption,
message authentication codes, pseudorandom functions etc... In
particular, very high speed applications shun digital signatures
in favor of message authentication codes,\footnote{In contrast
to digital signatures that have a public verification key and
secret signing key, \emph{message authentication codes} have a
single key for both verification and signing, and hence that key
must be kept private to maintain security.} even though the
latter sometime incur a significant cost in keeping shared
private keys among the entities involved (e.g., see \cite{tesla}
and the references therein).  The reason is that known
constructions of such schemes from one-way functions or other
unstructured primitives are quite \emph{inefficient}. This
problem already arises in \emph{one-time signatures}
\cite{Rabin78,Lamport79,Merkle87}, that are a relaxation of
digital signatures offering security only in the case that the
attacker observes  at most a single valid signature. The best
known constructions for this case require $\Omega(k)$
invocations of the one-way function (or even a random oracle) to
achieve $2^k$ security. In contrast, there are known
constructions of message authentication codes, private key
encryptions, and pseudorandom generators and functions that use
only $O(1)$ queries to a random oracle.

In this paper, we study the question of whether there exist more
efficient constructions of signature schemes from symmetric
primitives such as hash functions and block ciphers. We show to a
certain extent that the inefficiency of the known constructions is
\emph{inherent}.

\nnspace\SubSection{Our results}

We consider the efficiency of constructions of one-time
signatures using black boxes / oracles that model ideal
symmetric primitives: the random oracle, the random permutation
oracle, and the ideal cipher oracle (see
Section~\ref{sec:prelim} for definitions). We wish to study the
security of such constructions as a function of the number of
queries made to the oracle by the construction (i.e., by the
generation, signing, and verification algorithms). Of course, we
believe that one-time signatures exist and so there are in fact
signature schemes achieving super-polynomial security without
making any query to the oracle. Hence we restrict ourselves to
bounding the \emph{black-box} security of such schemes. We say
that a cryptographic scheme using oracle $\cO$ has
\emph{black-box security $S$} if for every $1 \leq T \leq S$, a
(potentially computationally unbounded) adversary that makes at
most $T$ queries to $\cO$ cannot break the scheme with
probability larger than  $T/S$ (see
Definition~\ref{def:sigsec}). Our main result is the following:

\begin{theorem} \label{thm:mainintro} Any one-time signature scheme
for $n$-bit messages using at most $q \leq n$ queries to a random
oracle has black-box security at most $2^{(1+o(1))q}$ where $o(1)$ goes to zero with $q$.
\end{theorem}

This is in contrast to other primitives such as message
authentication codes, collision resistant hash functions,
private-key encryption, and pseudorandom functions, that can all
be implemented using one or two queries to a random oracle with
black-box security that depends exponentially on the length of
these queries.  We note that Theorem~\ref{thm:mainintro} is
tight up to a constant factor in the number of queries, since a
simple modification of Lamport's scheme \cite{Lamport79} yields
$2^{(\alpha - o(1))q}$ black-box security, where $\alpha \sim
0.812$ is equal to $H(c)/(1+c)$, where $H$ is the Shannon
entropy function and  $c = (3-\sqrt{5})/2$ (see
Section~\ref{app:upperbound}). We also prove several extensions
of the main result:

\begin{description}

\nnspace\item[Other oracles.] Since our goal is to find out whether
signatures can be efficiently constructed from symmetric primitives,
it makes sense to study also other primitives than the random
oracle. Theorem~\ref{thm:mainintro}  extends (with a loss of a
constant factor in the number of queries) to the \emph{ideal cipher
oracle} and \emph{random permutation oracle} that are also sometimes
used to model the idealized security of symmetric primitives such as
block ciphers and one-way permutations.

\nnspace\item[Implementing adversary in $\BPP^{\NP}$.]  The proof of
Theorem~\ref{thm:mainintro} shows that for every $q$-query one-time
signature scheme for $\bits^n$ from random oracle, there is an adversary that
breaks it with probability close to $ 1$ using at most $\poly(q)2^q$
queries. However, the \emph{running time} of this adversary can be
higher than that. This is inherent, as otherwise we would be proving
unconditionally the non-existence of one-time signature schemes.
However, we show that this adversary can be implemented in
probabilistic polynomial-time using an oracle to an $\NP$-complete
problem. Thus, similar to what Impagliazzo and Rudich
\cite{ImpagliazzoRu89} showed for key-exchange, if there were a
{more efficient} construction of signature schemes from random
oracles with a proof of security relying on the adversary's
efficiency, then this is also a proof that $\P\neq\NP$.

\nnspace\item[Imperfect completeness.]  While the standard
definition of signature schemes requires the verifier to accept
valid signatures with probability $1$, one can also consider relaxed
variants where the verifier has some small positive probability of
rejecting even valid signatures. We say that such signature schemes
satisfy \emph{imperfect completeness}. We can extend
Theorem~\ref{thm:mainintro} to this case, though to get an attack
succeeding with high probability we lose a quadratic factor in the
number of queries.

\nnspace\item[Efficiency of the verifier.]  Because the
signing and the verification algorithms are executed more
often than the key generation algorithm, it makes sense to
study their efficiency separately rather than just studying
the total number of queries. Although in the construction
for signature schemes that we will see later (see
Section~\ref{app:upperbound}), the signing algorithm asks
only one oracle query and the total number of queries is
optimal up to a constant factor, the question about the
efficiency of the verifier still remains. We show that
(keeping the number of signing queries fixed to one) there
is a tradeoff between the number of queries asked by the
verification algorithm and the total number of queries,
conditioned on getting certain black-box security.

\nnspace\item[Black-box constructions.] As mentioned above,
all the symmetric primitives can be constructed from random
oracle, random permutation oracle, or ideal cipher oracle by
only $O(1)$ queries and get exponential security over the
length of the queries. Therefore, our lower bounds on
signatures from ideal oracles yield as corollaries lower
bounds on the efficiency of signatures from symmetric
primitives when the construction is black box. This holds
even when the one-way permutation used in the construction
has $n/2$ hardcore bits. The latter answers a
question raised by \cite{GennaroGeKaTr05}. Our results reject the existence of black-box constructions unconditionally (similar
to \cite{HaitnerHoReSe07}, while the results of \cite{GennaroGeKaTr05} show the existence of one-way function as a consequence. We prove the strongest possible form of lower bound on the efficiency of black box constructions of signatures from symmetric primitives.  Namely, we show that black-box constructions  of signature schemes for $n$-bit messages based on exponentially hard symmetric primitives of security parameter $n$, need to make at least $\Omega(n)$ calls to the primitive.

\end{description}

\nnspace\paragraph{Note on the random oracle model.} Although
the \emph{random oracle model} \cite{BellareRo93} (and its
cousin the \emph{ideal cipher model}) is frequently used as an
idealization of the properties enjoyed by certain constructions
such as the SHA-1 hash function \cite{SHA1} and the AES block
cipher \cite{AES02}, it has drawn a lot of criticism as this
idealization is not generally justified \cite{CanettiGoHa98}.
However,  for the sake of \emph{lower bounds} (as is our concern
here) this idealization seems appropriate, as it is a clean way
to encapsulate all the attractive properties that could be
obtained by constructions such as SHA-1,AES, etc..

\nnspace\paragraph{Taxonomy of black-box reductions.} Reingold,
Trevisan and Vadhan \cite{ReingoldTrVa04} study various notions of
``black-boxness'' of security proofs in cryptography according to
whether a construction of a cryptographic tool based on an
underlying primitive uses this primitive as a black box, and whether
its security proof uses the adversary as a black box. Those
definitions are not in the oracle model that we are concerned here. They call a construction for primitive $A$ from primitive $B$ black-box, if the implementation of $A$ uses $B$ as a black box. The security reduction which converts an adversary  for the implementation of $A$ to an adversary for $B$ could have different levels of being black box \footnote{It could be fully black-box, semi black-box, or non-black-box, and if the implementation reduction is black box, the whole construction is called, (resp.) fully black-box, semi black-box, or weakly black-box.}.  However, in the oracle based constructions studied here, the implementation reduction is always forced to be black-box, and for the proof of security, there is no security measure defined for the primitive used (i.e. the oracle) to which we could reduce the security of our construction. One common way to prove security for oracle based constructions is to rely on the statistical properties of the oracle and show that any (even computationally unbounded) adversary  breaking the implementation needs to ask \emph{many} queries from the oracle.
This gives a quantitative security guarantee and is called a \emph{black-box} proof of security in the oracle model.
A \emph{non-black-box} proof of security in this model, is a proof showing that any adversary who runs in time
$\poly(n, T)$ where $n$ is the input length and $T$ the number of oracle queries it asks, needs to ask many queries from the oracle.
\Mnote{I think this is the right way to call them, because it is meaningless to deprive the adversary from the oracle.
It was meaningful in the standard setting because the adversary can have the efficiently implemented used primitive in its belly.
We are exploiting the efficiency of the adversary to the maximum level that it could be exploited.}
In this work, we give
a lower bound on the number of queries needed to get  \emph{black-box} security $S$ for one-time signatures in various ideal
oracle models, and also show that if $\P=\NP$, then this bound holds for
\emph{non-black-box} proofs of security as well.
We note that if one-way functions exist, then there do exist
constructions making no query to the random oracle with
super-polynomial non-black-box security. \Mnote{because, if Eve wants to break the system, she should run in time more than poly(n), and then she needs to ask many (artificial) queries from the oracle to be able to run in that time, and that is what we need show in non-BB proofs.} As we mentioned before,
our  lower bounds in the ideal oracle models yield some lower bounds
on the efficiency of one-time signatures from symmetric primitives
in the standard model of \cite{ReingoldTrVa04}. We also note that
there do exist cryptographic constructions that use the primitive
\cite{GoldreichMiWi91,GoldreichMiWi87} or the adversary
\cite{Barak01} in a non-black-box way, but at the moment all of the
known highly efficient cryptographic constructions (e.g., those used
in practice) are black box, in the sense that if they use a generic
underlying  primitive (i.e., not based on specific problems such as
factoring) then it's used as a black-box and if they have a proof of
security then the proof treats the adversary as a black box.

\nnspace\SubSection{Prior work}

To the best of our knowledge, this is the first lower bound on the
number of random oracle queries needed to construct signature
schemes. Starting with the seminal paper of Impagliazzo and Rudich
\cite{ImpagliazzoRu89}, that showed that there is no construction of
a key exchange protocol from a random oracle with super-polynomial black-box security, and therefore rejecting black-box constructions of key exchange protocols from one-way function,  several works have investigated the \emph{existence}
of black-box constructions reducing one kind of cryptographic
scheme to another. However, only few works studied the
\emph{efficiency} of such constructions \cite{KimSiTe99,
GennaroGeKaTr05}. Of these, the most relevant is the paper by
Gennaro, Gertner, Katz, and Trevisan \cite{GennaroGeKaTr05}. They
considered the efficiency of basing various cryptographic primitives
on one-way permutations (OWP) secure against $S$-sized circuits, and
proved that to achieve super-polynomial security \textbf{(1)}
pseudorandom generators with $\ell$ bits of stretch require
$\Omega(\ell/\log S)$ invocations of the OWP, \textbf{(2)} universal
one-way hash functions compressing their input by $\ell$ bits
require $\Omega(\ell/\log S)$ invocations, \textbf{(3)} private key
encryption schemes for messages of length $n$ with key length $k$
require $\Omega((n-k)/ \log S)$ invocations, and (most relevant for
us) \textbf{(4)} one-time signature schemes for  $n$-bit messages
require $\Omega(n/\log S)$ invocations.\footnote{Otherwise, we can construct a one-way function directly.}

However, the one-way permutation oracle used by
\cite{GennaroGeKaTr05} was very far from being a random
oracle.\footnote{They considered an oracle that applies a random
permutation on the first $t$ bits of its $n$-bit input, for $t
\ll n$, and leaves the rest of the $n-t$ bits unchanged. This is
a one-way permutation with $2^{\Omega(t)}$ security.} Indeed,
the applications \textbf{(1)}, \textbf{(2)}, and \textbf{(3)}
can be implemented using only a constant number of calls to a
random oracle, and correspondingly are considered to have
efficient practical implementations. Thus,
\cite{GennaroGeKaTr05} did not answer  the question of whether
signature schemes can be efficiently constructed from efficient
symmetric key primitives such as hash functions and block
ciphers. \Mnote{Did they ask it? I think by picking the first
$t$ bits at random, we probably get even a hash function with
security $2^{\Omega(t)}$, and they can get similar results of
that type.} It is this question that we are concerned with in
this paper. Thus, on a technical level our work is quite
different from [GGKT] (as we work with a random oracle and
cannot ``tamper'' with it to prove our lower bound) and in fact
is more similar to the techniques in the original work of
Impagliazzo and Rudich \cite{ImpagliazzoRu89}. We note that this
work partially answers a question of \cite{GennaroGeKaTr05}, as
it implies that any black-box construction of  one-time
signatures from one-way permutation $p : \bits^n \mapsto \bits^n
$ with even $n/2$ hard-core bits requires at least $\Omega(n)$
queries to the permutation.

Several works
\cite{Merkle87,EvenGoMi96,Vaudenay92,BleichenbacherMa94,BleichenbacherMa96}
considered generalizations of Lamport's one-time signature
scheme. Some of these achieve shorter keys and signatures,
although their relation between the number of queries and
security (up to a constant factor) is at most a constant factor
better than Lamport's scheme (as we show is inherent).

\nnspace\Section{Our techniques} \label{sec:techniques}

We now give a high level overview of the ideas behind the proof
of Theorem~\ref{thm:mainintro}. Our description ignores several
subtle issues, and the reader is referred to
Section~\ref{sec:main} for the full proof.  To understand the
proof of the lower bound,\footnote{We use the terms ``lower
bound'' and ``upper bound'' in their traditional
crypto/complexity meaning of negative results vs. positive
results. Of course one can view Theorem~\ref{thm:mainintro} as
either upper-bounding the security or lower-bounding the number
of queries.} it is instructive to review the known \emph{upper
bounds} and in particular the simple one-time signature scheme
of Lamport \cite{Lamport79}. To sign messages of length $n$ with
security parameter $\ell$ using a random oracle $\cO$ (that we
model as a random function from $\bits^{\ell}$ to
$\bits^{\ell}$) the scheme works as follows:

\begin{itemize}

\item Generate the public verification key $VK$ by choosing $2n$
random strings $\{x_i^b \}_{i\in[n],b\in\bits}$ in $\bits^{\ell}$
and setting $VK$ to be the sequence $\{ y_i^b
\}_{i\in[n],b\in\bits}$ for $y_i^b = \cO(x_i^b)$.

\nnspace \item To sign a message $\alpha \in \bits^n$, simply reveal
the preimages in the set $\{ x_i^b \}_{i\in[n],b\in\bits}$ that
correspond to the bits of $\alpha$. That is, the signature is
$x_1^{\alpha_1},\ldots,x_n^{\alpha_n}$.

\nnspace\item The verifier checks that indeed
$\cO(x_i^{\alpha_i})=y_i^{\alpha_i}$ for every $i\in [n]$.
\end{itemize}

\nnspace This scheme uses $3n$ queries. It can be shown that it
has $2^{\Omega(\ell)}$ security. Note that in this case the
security can be arbitrarily large \emph{independently} of the
number of queries. Indeed, note that Theorem~\ref{thm:mainintro}
requires that the number of queries $q$ is not larger than the
length of the messages to be signed. Lamport's scheme can be
easily modified to work for unbounded size messages by following
the well known ``hash-and-sign'' paradigm: first use the random
oracle to hash the message to length $k$, and then apply
Lamport's scheme to the hashed value. This will result in a
scheme with $3k+2$ queries and (by the birthday bound) $2^{k/2}$
black-box security (see Section~\ref{app:upperbound} for some
improvements). We see that now indeed the security is bounded by
$2^{O(q)}$ (where $q = 3k+2$ is the number of queries),
regardless of the length $\ell$ of the queries.

The above discussion shows that to prove
Theorem~\ref{thm:mainintro}, we will need to use the fact that
there is a large number of potential messages, which is indeed
what we do. Note that the reason that the hash-and-sign variant
of Lamport's scheme only achieves $2^{k/2}$ security is that if
a pair of messages $\alpha,\beta$ satisfies
$\cO_k(\alpha)=\cO_k(\beta)$ (where $\cO_k(x)$ denotes the first
$k$ bits of $\cO(x)$), then they have the same signature, and so
a signature for $\alpha$ allows an adversary to forge a
signature on $\beta$. We will try to generalize this observation
to arbitrary signature schemes. For every such scheme
$\mathcal{S}$ and two messages $\alpha,\beta$ (after fixing the
oracle and the randomness of the system), we will say that
``$\alpha$ is useful for $\beta$'' if they satisfy a certain
condition. Then (roughly speaking) we will prove that:
\textbf{(A)} if $\alpha$ is useful for $\beta$ then a signature
on $\alpha$ can be used to compute a signature on $\beta$ by
asking at most $ 2^{O(q)}$ oracle queries (where $q$ is the
total number of queries made by the scheme $\mathcal{S}$), and
\textbf{(B)} if $\alpha$ and $\beta$ are chosen at random from a
large enough space of messages, then $\alpha$ will be useful for
$\beta$ with probability at least $2^{-O(q)}$. Together
\textbf{(A)} and \textbf{(B)} imply that, as long as the space
of possible messages is large enough, then the black-box
security of $\mathcal{S}$ is bounded by $2^{O(q)}$, since the
adversary can find a useful pair of messages $\alpha,\beta$ with
probability $2^{-q}$, ask for a signature on $\alpha$ and use
that to forge a signature on $\beta$ by asking $2^{q}$
queries.\footnote{The actual adversary we'll show will operate
by asking $\poly(q)2^q$ queries, and it succeeds  with
probability
 almost $1$, see the proof of Theorem~\ref{thm:main}.}

\vspace{1ex} \noindent \textsc{Defining the usefulness condition.}
This proof strategy rests of course on the ability to find an
appropriate condition ``$\alpha$ is useful for $\beta$'' for every
one-time signature scheme $\mathcal{S}$. This is what we describe
now. For now, we will assume that only the key generation algorithm of
$\mathcal{S}$ is probabilistic, and that both the signing and
verification algorithms are deterministic.\footnote{We study the randomized verifier in Section~\ref{subsec:ImpComp}, but assuming that the signer  is deterministic is without loss of generality. That is because the key generator can give, through the secret key, a secret seed $s$ to the signer, and the signer would use $\cO(s,\alpha)$ as the randomness needed to sign the message $\alpha$.} For every fixed randomness for the
generation algorithm, fixed oracle, and a message $\alpha$, we define
$G,S_{\alpha}$ and $V_{\alpha}$ to be the sets of queries
(resp.) made by the generation, signing,
and verification algorithms where the last two are applied on the
message $\alpha$.

\nnspace\paragraph{First attempt.} Observe that in the hash-and-sign
variant of Lamport's scheme, $\alpha$ and  $\beta$ have the same
signature if $V_{\alpha} = V_{\beta}$. This motivates stipulating
for every signature scheme  that $\alpha$ is useful for $\beta$ if
$V_{\beta} \subseteq V_{\alpha}$. This definition satisfies
Property~\textbf{(A)} above: if we know all the queries that the
verifier will make on a signature of $\beta$, then finding a
signature that makes it accept can be done by an exponential-time
exhaustive search that does not make any oracle queries at all. The
problem is that it might not satisfy \textbf{(B)}: it's easy to make
the verifier ask, when verifying a signature for $\alpha$, a query
that uniquely depends on $\alpha$, thus ensuring $V_{\beta}
\nsubseteq V_{\alpha}$ for every distinct $\alpha,\beta$.

\nnspace\paragraph{Second attempt.} A natural intuition is that
verifier queries that do not correspond to queries made by the
generation algorithm are sort of ``irrelevant''--- after all, in
Lamport's scheme all the queries the verifier makes are a subset
of the queries made by the generation algorithm. Thus, we might
try to define  that $\alpha$ is useful for $\beta$ if $V_{\beta}
\cap G \subseteq V_{\alpha}$. Since $G$ has at most $q$ queries,
and so at most $2^q$ subsets, this definition satisfies
Property~\textbf{(B)} since if $\alpha$ and $\beta$ are randomly
chosen from a set of size $2^q$ then $\alpha$ will be useful for
$\beta$ with probability at least $2^{-2q}$. Unfortunately, it
does not satisfy Property~\textbf{(A)}: there is a signature
scheme for which every pair of messages $\alpha,\beta$ satisfies
this condition even when a signature for $\alpha$ cannot be used
to forge a signature on $\beta$.\footnote{Such an example can be
obtained by the variant of Lamport's scheme where each signer
uses the verification key $VK$ to sign a new verification key
$VK'$ (the randomness for which is part of the secret key), and
then signs the message using the secret key corresponding to
$VK'$. In this case $V_{\alpha} \cap G = V_{\beta} \cap G$ for
every pair $\alpha,\beta$, even if a signature on $\alpha$
cannot be used to compute a signature on $\beta$.}

\nnspace\paragraph{Our actual condition.} The condition we
actually use, roughly speaking, is that $\alpha$ is useful for
$\beta$ if
\begin{equation}
V_{\beta} \cap (G \cup S_{\alpha}) \subseteq V_{\alpha} \;\;. \label{eq:useful}
\end{equation}
Using Bollob{\'a}s's Inequality \cite{Bollobas65} (see the proof
of Claim~\ref{claim:T3}) it can be shown that the condition
(\ref{eq:useful})  satisfies Property~\textbf{(B)}. It's less
obvious why it satisfies Property~\textbf{(A)}--- to see this we
need to see how our adversary will operate. The high level
description of our attack is as follows:

\begin{enumerate}
\item \textbf{Input: Key Generation.} The adversary receives the verification key $VK$.

\item \textbf{Request Signature.} Choose $\alpha \neq \beta
    \gets_R \bits^n$ at random, and get $\sigma_{\alpha}$,
    the signature of $\alpha$.

\item \textbf{Learning Oracle Queries.} Run $\Ver(VK,
    \alpha, \sigma_{\alpha})$ to learn the set $V_{\alpha}$
    of oracle queries that it asks and their answers. (Later
    we will modify this step somewhat, and ask some more
    oracle queries.)

\item \textbf{Sampling a Possible Transcript.} Conditioned
    on knowing $VK, \sigma_{\alpha}$, and answers of
    $V_{\alpha}$, \emph{guess}: the value of $SK$, the sets
    $G$ and $S_{\alpha}$, and their answers. Let
    $\Tilde{SK}$, $\Tilde{G}$, and $\Tilde{S}_{\alpha}$ be
    the guesses.

\item \textbf{Forging.} Sign the message $\beta$ by using
    $\Tilde{SK}$ and sticking to the oracle answers guessed
    for queries in $\Tilde{G} \cup \Tilde{S}_{\alpha}$ to
    get $\sigma_{\beta}$. That is, if we wanted to ask a an
    oracle query in $\Tilde{G} \cup \Tilde{S}_{\alpha}$, use
    the guessed answer, and otherwise ask the real oracle
    $\cO$. Output $\sigma_{\beta}$.

\end{enumerate}

Note that the queries for which we might have guessed a wrong
answer are in the set $(\Tilde{G} \cup \Tilde{S}_{\alpha})
\backslash V_{\alpha}$, because we did the guesses conditioned
on knowing $V_{\alpha}$ and its answers. Suppose that during the
verification of $(\beta, \sigma_{\beta})$, none of these queries
is asked from the oracle (i.e. $V_{\beta} \cap (\Tilde{G} \cup
\Tilde{S}_{\alpha} ) \subset V_{\alpha}$). Then we can
\emph{pretend} that our guesses were correct. That is, because
the answers to different queries of random oracle are
independent, as far as the verifier is concerned our guesses
could be right, and hence by definition, the verification of
$(\beta, \sigma_{\beta})$ must accept with probability $1$.

The description of the attack above shows that a similar
condition to the condition (\ref{eq:useful}), namely
\begin{equation}
V_{\beta} \cap (\Tilde{G} \cup \Tilde{S}_{\alpha}) \subset
V_{\alpha} \;\;, \label{eq:useful2}
\end{equation} has Property~\textbf{(A)}. But condition (\ref{eq:useful2}) might
not have Property~\textbf{(B)}. We cope with this by ensuring
that the attacker has sufficient information so that
(essentially) whenever  (\ref{eq:useful}) happens,
(\ref{eq:useful2}) also happens. This is accomplished by
learning more oracle queries before making the guesses. Namely,
we learn all the queries that are in the set $\Tilde{G} \cup
\Tilde{S}_{\alpha}$  with some noticeable probability
(conditioned on what we know about them). We then use a careful
hybrid argument  (that involves the most technical part of the
proof) to show that after performing this learning, the
condition (\ref{eq:useful2}) occurs with probability at least as
large as the probability that (\ref{eq:useful2}) occurs (up to
some lower order terms). Thus our actual usefulness condition
will be (\ref{eq:useful2}), though for the complete definition
of the sets $\Tilde{G} , \Tilde{S}_{\alpha}$ involved in it, one
needs to go into the details of the proof of
Theorem~\ref{thm:main}).

\nnspace\Section{Preliminaries} \label{sec:prelim}

\subsection{Basic Probability Facts}

We recall some simple but useful well known facts and
definitions about random variables.

\begin{definition} \label{def:SDDef}
The \emph{statistical distance} of two finite random variables
$X, Y$, denoted by  $\SD(X,Y)$, is defined to be
$\frac{1}{2}\sum_{a} |\Pr[X = a] - \Pr[Y=a] |$.
\end{definition}

\begin{lemma} \label{lem:SDEventBound}
If $A, B$ are random variables, and the event $E$ is defined
over $\Supp(A) \cup \Supp(B)$ (where $\Supp(X)$ denotes the
support of the random variable $X$), then $\abs{\Pr[E(A)] -
\Pr[E(B)]} \leq \SD(A, B)$.
\end{lemma}

\begin{lemma} \label{lem:SDforward}
If the random variable $A'$ is a function of random variable
$A$, and the random variable $B'$ is a function of $B$, then
$\SD(A', B') \leq \SD(A, B)$.
\end{lemma}

\begin{lemma} \label{lem:SDBound} If the event $E$ is defined over the random variable $A$, and the event $D$ is defined over the random variable $B$, and we have $\SD(A \mid E, B \mid D) = 0$, then $\SD(A, B) \leq (\Pr[E] + \Pr[D]) /2 $.
\end{lemma}

By $U_n$ we mean the uniformly distributed random variable over $n$-bit strings.

\subsection{Signature Schemes in Oracle Models}

We define the notion of one-time signature schemes and their
black-box security. We specialize our definition to the case that
the signature schemes use an oracle $\cO$ that may also be chosen
from some probability distribution. We use the standard notation
$A^{\cO}(x)$ to denote the output of an algorithm $A$ on input $x$
with access to oracle $\cO$.

\nnspace\begin{definition} \label{def:sig} An oracle \emph{signature
scheme} (with perfect completeness) for $n$ bit messages is a
triple of  oracle algorithms
$(\Gen,\Sign,\Ver)$ (where $\Gen$ could be probabilistic) with the following property: for every oracle
$\cO$, if $(SK,VK)$ is a pair that is output by $\Gen^{\cO}(1^n)$
with positive probability, then for every $\alpha \in \bits^n$,
$\Ver^{\cO}(VK,\alpha,\Sign^{\cO}(SK,\alpha))=1$. We call $SK$ the \emph{signing key} and $VK$ the
\emph{verification key}.
\end{definition}

One can also make a relaxed requirement that the verification
algorithm only needs to accept valid signatures with probability
$0.9$ (where this probability is over the verifier's coins
only). We say that such relaxed signature schemes have
\emph{imperfect completeness}, and we will consider such schemes
in Section~\ref{subsec:ImpComp}.  If the oracle algorithms of
the Definition~\ref{def:sig} run in polynomial-time, then we
call the signature scheme \emph{efficient}.  Note that we
consider (not necessarily efficient) signature algorithms on a
finite set of messages. For upper bounds (i.e., positive
results) one would want uniform efficient algorithms that could
handle any size of message, but for a lower bound (i.e., a
negative result), this simpler definition will do.

So far, we did not say anything about the security. In the following
definition we specify the ``game'' in which the adversary
participates and tries to break the system and give a
quantitative measure for the security.

\nnspace\begin{definition} \label{def:sigsec} For every $S\in\N$,
the oracle signature scheme $(\Gen,\Sign,\Ver)$ is a \emph{one-time}
signature scheme with \emph{black-box security} $S$, if for every
message $\alpha \in \bits^n$, $1 \leq T \leq S$, and adversary algorithm
$A$ that makes at most $T$  queries to its oracle, $\Pr[
\Ver(VK,\alpha^*,\sigma^*)=1 \text{ where } (\alpha^*,\sigma^*) =
A^{\cO}(VK,\Sign^{\cO}(SK,\alpha)) \text{ and } \alpha^* \neq \alpha
] \leq \tfrac{T}{S}$, where $(SK,VK) = \Gen^{\cO}(1^n)$, and this
probability is over   the coins of all algorithms
($\Gen,\Sign,\Ver$, and $A$), and the choice of the oracle $\cO$.
\end{definition}

This is a slightly weaker definition of security than the standard
definition, since we are not allowing the adversary to choose the
message $\alpha$ based on the public key. However, this is again
fine for lower bounds (the known upper bounds do satisfy the
stronger definition). Also, some texts use $1/S$ (rather than $T/S$)
as the bound on the success probability. Security according to
either one of these definitions is always at most quadratically
related, but we feel Definition~\ref{def:sigsec} is more precise.

In a \emph{non-black-box} proof of security, the running time of
the adversary is utilized in order to prove the security of the
system:

\nnspace\begin{definition} \label{def:weaksigsec} For every
$S\in\N$, the oracle signature scheme $(\Gen,\Sign,\Ver)$ is a
one-time signature scheme with \emph{non-black-box} security $S$,
if for every message $\alpha \in \bits^n$, $T \leq S$, and adversary
algorithm $A_T$ that makes at most $T$ oracle queries and runs in
time $\poly(n,T)$, $\Pr[ \Ver(VK,\alpha^*,\sigma^*)=1 \text{ where }
(\alpha^*,\sigma^*) = A_T^{\cO}(VK,\Sign^{\cO}(SK,\alpha)) \text{
and } \alpha^* \neq \alpha ] \leq \tfrac{T}{S}$, where $(SK,VK) =
\Gen^{\cO}(1^n)$, and this probability is over the coins of all
algorithms ($\Gen,\Sign,\Ver$, and $A_T$), and the choice of the
oracle $\cO$.
\end{definition}

\nnspace\paragraph{Oracles.} In this work, as for the oracle
signature schemes,  we only use one of the following oracles:
\textbf{(1)} The \emph{random oracle} returns on input
$x\in\bits^n$ the value $f(x)$ where $f$ is a random function
from $\bits^n$ to $\bits^n$.\footnote{More generally, $f$ can be
a function from $n$ to $\ell(n)$ for some function
$\ell:\N\To\N$, but using standard padding arguments we may
assume $\ell(n)=n$.} \textbf{(2)} The \emph{random permutation
oracle} returns on input $x\in\bits^n$ the value $f(x)$ where
$f$ is a random permutation on $\bits^n$. \textbf{(3)} The
\emph{ideal cipher oracle} with message length $n$, returns on
input $(k,x,d)$ where $k\in\bits^*$, $x\in\bits^{n}$ and $d \in
\{ \mathsf{F},\mathsf{B}\}$, $f_k(x)$ if $d = \mathsf{F}$ and
$f_k^{-1}(x)$ if $d = \mathsf{B}$, where for every
$k\in\bits^*$, $f_k$ is a random permutation on $\bits^{n}$.
\Mnote{I changed the definition of ideal cipher a bit, to
arbitrary size keys, our proof works for this case.} These three
oracles are standard idealizations of (respectively) hash
functions, one-way permutations, and block ciphers (see also
Section~\ref{subsec:BBconst}).

\nnspace\Section{Proof of the main result} \label{sec:main}

\begin{theorem} \label{thm:main}
Let $(\Gen, \Sign, \Ver)$ be a one-time oracle signature scheme
(with perfect completeness) in random oracle model for the space
of messages $\mathcal{M}$ in which the total number of oracle
queries asked by $\Gen, \Sign$, and $\Ver$ is at most $q$, and
$|\mathcal{M}| \geq \frac{{q \choose {q/2}}}{\lambda}$. Then
there is a (computationally unbounded) adversary which asks at
most $ O(\frac{q^2 {q \choose {q/2}} }{\lambda \delta^2}) =
O(\frac{q^{1.5} 2^q}{\lambda  \delta ^2})$ oracle queries and
breaks the scheme with probability $1- (\lambda + \delta)$. This
probability is over the randomness of the oracle as well as the
coin tosses of the key generation algorithm and the adversary.
\end{theorem}

\noindent Theorem~\ref{thm:main} implies
Theorem~\ref{thm:mainintro} via the following corollary:

\begin{corollary} \label{cor:main}
Let $(\Gen, \Sign, \Ver)$ be a one-time oracle signature for the messages $\mathcal{M} = \bits^n$ in the random oracle model in which the total queries asked by the scheme is at most $q$ where $q \leq n$, then there is an adversary asking $2^{(1+o(1))q}$ queries breaking the scheme with probability at least $1-o(1)$
and  at least $0.49$ for any $q \geq 1$.
\end{corollary}

\begin{proof} Let $\delta = \lambda = {q \choose q/2} / 2^q = \theta(q^{-1/2}) =  o(1)$, so  we have
$|\mathcal{M}| = 2^n \geq 2^q = {q \choose q/2} / \lambda$. Therefore we get an adversary asking
$O(q^{3.5} {q \choose q/2}) = O(q^{3} 2^q) = 2^{(1+o(1))q}$ queries breaking the scheme with probability $1-o(1)$. Thus the
 black-box security of the scheme  is at most by $\frac{2^{(1+o(1))q}}{1-o(1)} = 2^{(1+o(1))q}$.
For any $q \geq 1$, $\lambda$ can be as small as ${1 \choose 0}/ 2^1 = 1/2$, and by taking $\delta = 0.01$  the success
probability will be at least $0.49$.
\end{proof}

We now turn to proving Theorem~\ref{thm:main}. Let
$(\Gen,\Sign,\Ver)$ be as in the theorem's statement. We assume
that only $\Gen$ is probabilistic, and $\Sign$ and $\Ver$ are
deterministic. We also assume that all the oracle queries are of
length $\ell$. Since we assume the signature has perfect
completeness, these assumptions can be easily shown to be
without loss of generality. (In the case of imperfect
completeness the verifier algorithm is inherently probabilistic;
this case is studied in Section~\ref{subsec:ImpComp}.) We will
show an adversary that breaks the signature system with
probability $1-(\lambda + O(\delta))$, which implies
Theorem~\ref{thm:main} by simply changing $\delta$ to $\delta/c$
for some constant $c$.

\vspace{1ex} \noindent \textsc{The adversary's algorithm.}
Our adversary $\Adv$ will operate
as follows:

\begin{description}

\vspace{-1ex}\item[Input: Key generation.] The adversary receives a
verification key $VK$, where $(VK,SK) = \Gen(1^n)$.

\vspace{-1ex}\item[Step 1: Request signature.] Let
$\beta_0,\ldots,\beta_{N-1}$ denote the first $N = \frac{{q \choose q/2}}{\lambda}$ distinct messages
(in lexicographic order) in $\mathcal{M}$. Let
$\alpha_0,\ldots,\alpha_{N-1}$ be a random permutation of
$\beta_0,\ldots,\beta_{N-1}$. $\Adv$ asks for a signature on
$\alpha_0$ and verifies it (note that $\alpha_0$ is chosen
independently of the public key). We denote the obtained signature
by $\sigma_0$, and we denote by $T_0$ the \emph{transcript} of the algorithms run so far, which includes the random tape of the key generation algorithm,
all the queries made by the key generation, signing, and verification
algorithms, and the answers to these queries. So $T_0$ completely describes the running of the algorithms so far. (Note that $\Adv$ only
has partial information on $T_0$.)

\vspace{-1ex}\item[Step 2: Learning query/answer pairs.] We denote
by $L_0$ the information that $\Adv$ currently has on the oracle
$\cO$ and the randomness of the generation algorithm: that is, $L_0$ consists of $VK,\sigma_0$ and the
queries made by the verifying algorithm $\Ver$ on input
$VK,\sigma_0$, along with the answers to these queries. Let $\e =
\frac{\delta}{qN}$, and $M = \frac{q}{\e\delta} = \frac{q^2 N}{\delta^2}$. For $i=1,\ldots,M$, do the following:

\begin{enumerate}

\item Let $\mathbf{D}_{i-1}$ be the distribution of $T_0$, the transcript of the first step, conditioned on only knowing
$L_{i-1}$.

\item We let $Q(L_{i-1})$ denote the queries appearing in $L_{i-1}$.
If there exists a string $x \in \bits^{\ell} \setminus Q(L_{i-1})$
that is queried with probability at least $\e$ in $\mathbf{D}_i$,
then $\Adv$ lets $L_i$ be $L_{i-1}$ concatenated with the
query/answer pair $(x_i,\cO(x_i))$, where $x_i$ is the
lexicographically first such string. Otherwise, $L_i = L_{i-1}$.

\end{enumerate}

\vspace{-1ex}\item[Step 3: Sampling a possible transcript.] $\Adv$
generates a random transcript $\Tilde{T}_0$ according to the
distribution $\mathbf{D}_M$. Note that $\Tilde{T}_0$ also determines
a secret signing key, which we denote by $\Tilde{SK}$ ($\Tilde{SK}$
may or may not equal the ``true'' signing key $SK$). $\Tilde{T}_0$
may also determine some query/answer pairs that were not in $L_M$,
and hence may not agree with the the actual answers of the ``true''
oracle $\cO$. We denote by $\Tilde{\cO}$ the oracle that on input
$x$, if $x$ appears as a query in $\Tilde{T}_0$ then
$\Tilde{\cO}(x)$ outputs the corresponding  answer, and otherwise
$\Tilde{\cO}(x)=\cO(x)$.

\vspace{-1ex}\item[Step 4: Forging.] For every $j=1,\ldots,N-1$,
$\Adv$ uses $\Tilde{SK}$ and the oracle $\Tilde{\cO}$ to compute a
signature on the message $\alpha_j$, which it then tries to verify
this time using $VK$ and the ``true'' oracle $\cO$. $\Adv$ outputs
the first signature that passes verification.

\end{description}

\noindent\textsc{Analysis.} The number of queries asked during the
attack is at most $M+ qN = \frac{q^2 N}{\delta^2} + qN  \leq \frac{2 q^2 N}{\delta^2} = O(\frac{q^2 {q \choose {q/2}} }{\lambda \delta^2})$.  To
analyze the success probability of $\Adv$ we will prove the
following lemma:

\begin{lemma} \label{lem:main} For every $j \in [0..N-1]$,
let $V_j$ denote the set of queries made by $\Adv$ when verifying
the signature on $\alpha_j$. Let $\Tilde{G}$ and $\Tilde{S}_0$ be
the sets of queries made by the generation and signing algorithms
according to the transcript $\Tilde{T}_0$. For every $j\geq 1$, let
$E_j$ be the event that $V_j \cap (\Tilde{G} \cup \Tilde{S}_0)
\subseteq V_0$. Then,
\[
\Pr[ \cup_{j\in[1..N-1]} E_j ] = 1 - (\lambda + 2\delta)  \mper
\]
\end{lemma}

Note that the event $E_j$  corresponds to the condition that
``$\alpha_0$ is useful for $\alpha_j$'' described in
Section~\ref{sec:techniques}. Lemma~\ref{lem:main} implies
Theorem~\ref{thm:main} since if the event $E_j$ holds then when
verifying the signature for $\alpha_j$, the verifier never asks a
query on which the oracles $\cO$ and $\Tilde{\cO}$ differ (these
oracles can differ only on queries in $(\Tilde{G} \cup \Tilde{S}_0)
\setminus V_0$). But if the verifier uses the same oracle
$\Tilde{\cO}$ used by the generation and signing algorithm, then by
the definition of a signature scheme, it must accept the signature.

\nnspace\SubSection{Proof of Lemma~\ref{lem:main}}

It turns out that using known combinatorial techniques, one can
show that $\cup_j E_j$ holds with high probability if all
signatures and verifications were to use  the ``true'' oracle
$\cO$ and signing key $SK$ (as opposed to $\Tilde{O}$ and
$\Tilde{SK}$). The idea behind the proof is to show this holds
in our case using a hybrid argument. Specifically, we define
four distributions $\Hyb^0,\Hyb^1,\Hyb^2,\Hyb^3$, where $\Hyb^0$
corresponds to $\Tilde{T}_0$ joint with all the oracle
queries/answers that the adversary gets during the signing and
verification algorithms on $\alpha_j$ for $j \geq 1$ (we call
this information the \emph{transcript} of the experiment), and
$\Hyb^3$ corresponds to $T_0$ (the real transcript of the first
step) joint with the rest of the system's transcript if we  use
the ``true'' oracle and signing key (so the adversary is not
doing anything in generating $\Hyb^3$). \Mnote{I think the way
we had used the term "transcript" here was vague, specially for
$H^0$.} We will prove the lemma by showing that the probability
of $\cup_j E_j$ is almost the same in all these four
distributions.

\nnspace\paragraph{Definition of hybrid distributions.} The  four
hybrid distributions $\Hyb^0,..,\Hyb^3$ are defined as follows:

\begin{description}

\item[$\Hyb^0$:] This is the distribution of
$\Tilde{T}_0,T_1,\ldots,T_{N-1}$, where $\Tilde{T}_0$ denotes the
transcript sampled by $\Adv$ in Step~3, while $T_j$ (for $j\geq 1$)
denotes the transcript of the $j^{\tth}$ signature (i.e., the
queries and answers of the signing and verification algorithms  on
$\alpha_j$) as generated by $\Adv$ in Step~4. Note that $T_0$ and  $\Tilde{T}_0$ describe also the running of the key generation while $T_j$ for $j \geq 1$ do not.

\vspace{-1ex}\item[$\Hyb^1$:] This is the same distribution as
$\Hyb^0$, except that now in Step~4 of the attack, the adversary
uses the modified oracle $\Tilde{\cO}$ for both signing and
verifying the signatures on $\alpha_1,\ldots,\alpha_{N-1}$ (recall
that in $\Hyb^0$ the oracle $\Tilde{\cO}$ is only used for signing).

\vspace{-1ex}\item[$\Hyb^2$:] This is the same distribution as
$\Hyb^1$, except that we make a slight modification in the
definition of $\Tilde{\cO}$: for every query $x$ that was asked by
the generation, signing, and verification algorithms in the Input step and Step~1
(i.e., for every query in $T_0$), we answer with $\cO(x)$
\emph{only} if $x$ also appears in $L_M$. Otherwise, we answer this
query with a completely random value. Note that all the queries of the verification are in $L_0$ and so in $L_M$ as well. In other words, $\Tilde{\cO}$ agrees with $\cO$ on all the queries that $\Adv$ has asked from $\cO$ till the end of Step~2, and all the others are answered completely at random.

\vspace{-1ex}\item[$\Hyb^3$:] This is the same distribution as the previous ones, with the difference that $\Tilde{T}_0$ is chosen equal to $T_0$ (and so, there is no point in neither  Step~2 of the attack nor defining $\Tilde{\cO}$ anymore).
In other words, this is the transcript (randomness
and all query/answer pairs) of the following experiment:
\textbf{(1)} Generate signing and verification keys $(SK,VK)$ using
a random oracle $\cO$ \textbf{(2)} for $j=0\ldots N-1$, sign
$\alpha_j$ and verify the signature using $SK,VK$ and $\cO$.
\end{description}

\noindent Note that the hybrid distributions $\Hyb^i$ are over the
coin tosses of  the oracle, the key generation algorithm, and the
adversary. Lemma~\ref{lem:main} follows immediately from the
following claims:

\torestate{Claim}{claim:T0T1}{ $\Pr_{\Hyb^0}[ \cup_{j\geq 1} E_j ] =
\Pr_{\Hyb^1}[ \cup_{j\geq 1} E_j ]$. }

\torestate{Claim}{claim:T1T2}{$\SD(\Hyb^1, \Hyb^2) \leq
2\delta$.  Thus, $\Pr_{\Hyb^1}[ \cup_{j\geq 1} E_j ] \geq
\Pr_{\Hyb^2}[ \cup_{j\geq 1} E_j ] - 2\delta$.}

\torestate{Claim}{claim:T2T3}{$\Hyb^2 \equiv \Hyb^3$. Thus,
$\Pr_{\Hyb^2}[ \cup_{j\geq 1} E_j ] = \Pr_{\Hyb^3}[ \cup_{j\geq 1}
E_j ]$.}

\torestate{Claim}{claim:T3}{$\Pr_{\Hyb^3}[ \cup_{j\geq 1} E_j ] \geq
1-\lambda$.}

\nnspace\SubSection{Proof of Claims~\ref{claim:T0T1} to
\ref{claim:T3}}

We now complete the proof of Lemma~\ref{lem:main} by proving
Claims~\ref{claim:T0T1} to \ref{claim:T3}.

\restate{claim:T0T1}

\begin{proof}

Suppose we sample the hybrid distributions $\Hyb^0$ and $\Hyb^1$ using the same oracle $\cO$, same randomness for key generation, and the same randomness for the adversary. Then it is easy to see that for any $j$, the event $E_j$ holds for $\Hyb^0$ iff it holds for $\Hyb^1$ and so is the event $\cup_{j \geq 1} E_j$. This shows that the probability of
$\cup_{j \geq 1} E_j$ happening in both distributions is the same.
\end{proof}

\restate{claim:T1T2}

\begin{proof}
Let $B$ be the event that $\Adv$ asks a query in $Q(T_0) \setminus
Q(L_M)$,  where  $Q(T_0)$ denotes the queries in the
transcript $T_0$. It is easy to see that conditioned on $B$ doesn't happen  $\Hyb^1$
and $\Hyb^2$ are identically distributed. That is because if we use the same randomness for key generation, oracle and the adversary in the sampling of $\Hyb^1$ and $\Hyb^2$, conditioned on $B$ not happening (in both of them), the value of $\Hyb^1$ and $\Hyb^2$ is equal. In particular it shows that the probability of
of $B$  is the same in both distributions. Therefore the statistical
distance between $\Hyb^1$ and $\Hyb^2$ is bounded by the probability
of $B$.  In the following, we show that $\Pr_{\Hyb^2}[B] \leq 2\delta$. In the the following all the probabilities will be in the experiment for $\Hyb^2$.

Let $\e,\delta$ and $M$ be as in Step~2 of $\Adv$:  $\e =
\frac{\delta}{q N}$, and $M = \frac{q}{\e\delta}$. We start by showing:
$\Pr [ C ] \leq \delta$ where the event $C$ is defined as $$C \colon \exists x \not\in Q(L_M) \text{ that is obtained in }
\mathbf{D}_M \text{ with prob } \geq \e$$
and $\mathbf{D}_i$ is defined, as in Step~2 of $\Adv$ to be the
distribution of the transcript of the first signature conditioned on
the information in $L_i$.

\begin{quote}\textsc{Proof of $\Pr[C] \leq \delta$.}  For every possible
query $x$ to the random oracle, let $q_x$ denote the probability,
taken over both the random oracle and the randomness used by $\Gen$
and $\Adv$, that $x$ is queried when generating a key and then
signing and verifying $\alpha_0$.  Then $\sum_x q_x \leq q$
\textbf{(*)} since this sum is the expected number of queries in
this process. Let $p_x$ denote the probability that $x$ is learned
at some iteration of  Step~2. Then, $q_x \geq \e p_x$ \textbf{(**)}.
Indeed, if $A_i$ is the event that $x$ is learned at the $i^{\tth}$
iteration, then since these events are disjoint $q_x = \Pr[ x \text{
is queried} ] \geq \sum_{i=1}^M \Pr[x \text{ is queried} \mid
A_i]\Pr[A_i]$. But by definition of the learning process , $\Pr[x
\text{ is queried} \mid A_i] \geq \e$ and hence $q_x \geq \e
\sum_{i=1}^M \Pr[A_i] = \e p_x$. But the event $C$ only occurs if $M$ distinct queries are learned
in Step~2. Hence, if it happens with probability more than $\delta$
then the expected number of queries learned, which is $\sum_x p_x$,
is larger than $\delta M$. Yet combining \textbf{(*)} and
\textbf{(**)}, we get that $\delta M < \sum_x p_x \leq \sum_x q_x  /
\e \leq q/\e$, contradicting the fact that $M = q/(\e\delta)$. \qed
\end{quote}

Now we will show that $\Pr[B \mid \neg C] \leq \delta$, and it means that $\Pr[B] \geq \Pr[\neg C] \Pr[B \mid \neg C] \geq (1-\delta)^2   > 1-2\delta$.
Note that $\Adv$ makes all its operations in Step~4 based solely on the
information in $L_M$, and the answers chosen for queries $Q(T_0) \backslash Q(L_M)$ does not affect it (because even if queries
in $Q(T_0) \backslash Q(L_M)$ are asked by $\Adv$, they will be
answered at random). So, it means that  the value of $\Hyb^2$ is independent of $T_0$, conditioned on knowing $L_M$.
Thus, instead of thinking of $T_0$ being chosen first, then $L_M$
computed and then all queries of Step~4 being performed, we can
think of $L_M$ being chosen first, then $\Adv$ runs Step~4 based on
$L_M$ to sample $\Hyb^2$, and then $T_0$ is chosen conditioned on $L_M$ and $\Hyb^2$. But because of the independence of $T_0$ and $\Hyb^2$ conditioned on $L_M$,
the distribution of $T_0$ conditioned on $L_M$ and $\Hyb^2$ is that conditioned on only $L_M$ which has the distribution
$\mathbf{D}_M$. Now assume that $L_M$ makes the event $\neg C$ happen (note that $C$ is defined by $L_M$.).  Since at most $qN$ queries are made in
Step~4, and $C$ has not happened, when $T_0$
is chosen from $\mathbf{D}_M$, the probability that $Q(T_0)
\setminus Q(L_M)$ contains one of these queries is at most $\e q N
= \delta$. Therefore we get $\Pr[B \mid \neg C] \leq \delta$, and $\Pr[B] < 2\delta$.
\end{proof}

\restate{claim:T2T3}
\begin{proof}
In the sampling of $\Hyb^3$ we can think of $L_M$ being chosen first (although not needed), and then $T_0$ being chosen conditioned on $L_M$ (i.e., from
the distribution $\mathbf{D}_M$), and then
Step~4 of the experiment is done while any query in $Q(L_M) \cup Q(T_0)$ is answered according to $L_M, T_0$, and any
other query is answered randomly. (That is we sample $L_M$ and $T_0$ in the reverse
order.)
The point is that during the sampling process of $\Hyb^2$ we are also doing exactly the same thing. Again, we sample $L_M$ first. Then $\Tilde{T}_0$ is
chosen from the distribution $\mathbf{D}_M$. Then Step~4 is done while  any query in $Q(L_M) \cup Q(\Tilde{T}_0)$ is answered
according to $L_M, \Tilde{T}_0$, and all other queries
(even the ones in $Q(T_0) \backslash Q(L_M)$) are answered randomly. Therefore $\Hyb^2$ and $\Hyb^3$ have the same distribution.
\end{proof}

\restate{claim:T3}

\begin{proof} We will prove that this holds for every fixed oracle and
randomness of all parties, as long as the permutation
$\alpha_0,\ldots,\alpha_{N-1}$ is chosen at random. For every fixing
of the oracle and randomness and $j\in[0..N-1]$, let $U_j = G \cup S_{\beta_j}$ denote
the set of queries made by either the key generation algorithm or the
signing algorithm for message
$\beta_j$, and let $V_j$ be the set of queries made by the
verification algorithm while verifying this signature. The proof
will follow from this fact:

\vspace{1ex}\noindent\textsc{Combinatorial Lemma:} If
$U_1,\ldots,U_K,V_1,\ldots,V_K$ are subsets of some universe
satisfying $|U_i|+|V_i| \leq q$ and $U_i \cap V_j \nsubseteq V_i$
for every $i\neq j$ then $K \leq \binom{q}{q/2}$.

\vspace{1ex} The Combinatorial Lemma immediately implies
Claim~\ref{claim:T3}. Indeed, for every $i,j$ with $i \neq j$,
define the event $E_{i,j}$ to hold if $U_i \cap V_j \subseteq V_i$.
Then, there must be at least $N - \binom{q}{q/2} = N(1-\lambda)$ number of $i$'s
(i.e., $1-\lambda$ fraction of them) such that $E_{i,j}$ holds for some $j$ (otherwise we could remove
all such $i$'s and obtain a larger than $\binom{q}{q/2}$-sized family
contradicting the combinatorial Lemma). But, if we choose a
permutation $\alpha_0,\ldots,\alpha_{N-1}$ such that $\alpha_0 =
\beta_i$ for such an $i$ then the event $\cup_j E_j$ holds.

Thus, all that is left is to prove is the combinatorial lemma. It
essentially follows from Bollob{\'a}s's Inequality \cite{Bollobas65},
but we repeat the argument here. Assume for the sake of
contradiction that there is a family $U_1,\ldots,U_K,V_1,\ldots,V_K$
satisfying conditions of the lemma with $K>\binom{q}{q/2}$. First,
we can remove any elements from $U_i$ that are also in $V_i$, since
it will not hurt any of the conditions. It means that now we have: for every $i,j$, $U_i \cap V_j = \emptyset$ iff $i=j$. Now,
take a random ordering of the universe $W = \bigcup_i (U_i \cup
V_i)$, and let $A_i$ be the event that all the members of $U_i$
occur before the members of $V_i$ in this order. The probability of
$A_i$ is $\tfrac{|U_i|!|V_i|!}{(|U_i|+|V_i|)!} =
1/\binom{|U_i|+|V_i|}{|V_i|} \geq 1/\binom{q}{|V_i|} \geq
1/\binom{q}{q/2}$. Hence if $K > \binom{q}{q/2}$, there is a positive
probability that both $A_i$ and $A_j$ hold for some $i \neq j$. But
it is not hard to see that in that case, either $U_i$ and $V_j$ are
disjoint or $U_j$ and $V_i$ are disjoint, contradicting our
hypothesis.
\end{proof}

\nnspace\Section{A One-Time Signature Scheme}
\label{app:upperbound}

The following Theorem shows that Theorem~\ref{thm:mainintro} is
tight up to a constant factor in the number of queries.

\begin{theorem} \label{thm:LamChanged}
There is a one-time signature scheme
$(\Gen,\Sign,\Ver)$ for messages $\bits^*$, using a total of $q$ queries to a random oracle
that has security $2^{(0.812 - o(1))q}$, where $o(1)$ is a term
tending to $0$ with $q$.
\end{theorem}

\begin{proof}
The scheme is basically Lamport's Scheme \cite{Lamport79} with two
changes: \textbf{(1)} we use a more efficient anti-chain (family of
incomparable sets) than Lamport's scheme (a well-known optimization)
and \textbf{(2)} we use a secret ``salt'' value for the hash
function to prevent a birthday attack.

\vspace{1ex}\noindent\textsc{\bf The Scheme Description.} Let $c =
(3-\sqrt 5)/2$ and $k$ be such that $(1+c)k+4=q$.

\begin{itemize}

\item Generate the keys by choosing $k$ random strings
    $x_{i} \in \bits^{q+i}$ for $0 \leq i \leq k-1$, and an additional random
    string $z \in \bits^{2q}$. \footnote{If we choose all of them from $\bits^q$ the scheme is still as secure as we claimed, but now
    the analysis is simpler.} The secret key consists of these
    values, and the public key is
    $\cO(x_1),\ldots,\cO(x_k),\cO(z)$.

\item Let $h(\alpha)$ be the
    first $\log \binom{k}{ck}$ bits of
    $\cO(z,\alpha)$, which we identify with a $ck$-sized
    subset of ${0, \dots, k-1}$. The signature of $\alpha$ consists of $\{ x_i \}_{i
    \in h(\alpha)}$ and the string $z$.

\item To verify a signature, we first verify that $O(z)$ is equal to its alleged value,
then we ask $\cO(z,\alpha)$ to know $h(\alpha)$, and then we ask $ck$ more queries to check that the released strings are indeed
    preimages of the corresponding entries of the public key indexed by $h(\alpha)$.

\end{itemize}

The number of queries is $q=(1+c)k+4$, while, as we will see, the security is at
least $\Omega(\binom{k}{ck}) = 2^{(H(c)-o(1))k} =
2^{\frac{H(c)-o(1)}{1+c}q} > 2^{(0.812-o(1))q}$ where $H(\cdot)$ is
the Shannon entropy function.

Let $T$ be the total number of oracle queries asked by the adversary
and $\alpha \neq \beta$ be  (in order) the message for which she asks a
signature and the message for which she tries to forge a signature. We assume without loss of generality
that $T < 2^{q-1}$, because $2^{q-1}\gg \binom{k}{ck}$.
We divide the winning cases for the adversary into three cases:

\begin{enumerate}
\item The adversary chooses some $z' \in \bits^{2q}, z' \neq z$ such that $\cO(z) = \cO(z')$, alleged to be the real $z$ in the signature of $\beta$.
\item The adversary uses the real $z$ in the signature of $\beta$ and $h(\alpha) = h(\beta)$.
\item The adversary uses the real $z$ in the signature of $\beta$ and $h(\alpha) \neq h(\beta)$.
\end{enumerate}

We will show that the probability that the adversary wins
conditioned on being in case 3 is  at most $O(T/\binom{k}{ck})$,
and the probability that either case 1 or case 2 happens at all
is also at most $O(T/\binom{k}{ck})$. So, the total probability
of winning for the adversary will be at most
$O(T/\binom{k}{ck})$ as well.

In case 1, even if we reveal $z$ to the adversary in the first
place ($x_i$'s are irrelevant), she has the chance of at most
$(1+T)/2^q$ to find some $z'\neq z$ such that $\cO(z) =
\cO(z')$. That is because she gets to know at most $T$ oracle
query/answer pairs (other than $\langle z, \cO(z)\rangle$), and
the probability that she gets $\cO(z)$ in one of them is at most
$T/2^q$. If she does not see $\cO(z)$ as an oracle answer, she
needs to guess $z'$ blindly which succeeds with probability at
most $1/2^q$.

In the case 2, we reveal all $x_i$'s to the adversary at the beginning, although they are indeed irrelevant to finding a pair
$\alpha \neq \beta$ such that $h(\alpha) = h(\beta)$ (because they are of length $< 2q$).
Before the adversary gives us $\alpha$, it asks at most $T$ queries of length $2q$.
So, the probability
that she gets some $z' \in \bits^{2q}$ such that $\cO(z') = \cO(z)$ is at most $T/2^{2q} = o(T/\binom{k}{ck})$.
Let assume that this has not happened. So, we can pretend that when
we receive $\alpha$, the value of $z$ is chosen at random different from the members of $\bits^{2q}$ that are asked from the oracle by $\Adv$.
Thus, the probability that any adversary's query so far with length more than $2q$ has the prefix $z$ will be at most
$T/(2^q-T) < T/2^{q-1} = O(T/\binom{k}{ck})$. It means that with probability $1-O(T/\binom{k}{ck})$, so far were no query
asked from the oracle which has $z$ as prefix. Assuming this is the case, when we ask the query
$(z,\alpha)$ from the oracle, $h(\alpha)$ is chosen uniformly at random from $\bits^{\log \binom{k}{ck}}$.
Hence, if the adversary asks $T$ more oracle queries of the
form $(z, \gamma)$ where $\gamma \neq \alpha$, one of them will give $h(\gamma) = h(\alpha)$ with probability at
most $T/\binom{k}{ck}$, and if it does not happen for any of them, a blind
guess $\beta$ by the adversary will give $h(\alpha) = h(\beta)$ with probability $1/\binom{k}{ck}$.
So, the probability of getting $\alpha \neq \beta$, $h(\alpha) = h(\beta)$ for the adversary is at most $O(T/\binom{k}{ck})$.

In the case 3, there always is  some $i \in h(\beta) \backslash h(\alpha)$. We choose the smallest  such $i$, call it $i_0$,
and change the  game slightly by revealing $z$ to $\Adv$ from the beginning and revealing all $x_j$'s for $j \neq i_0$ to the her
after she gives us $\beta$.
It only might increases her chance of success (although they are irrelevant because they have different length). For any fixed
$i \in {0, \dots, k-1}$, we show
that the probability of the adversary to find a preimage for $\cO(x_i)$
conditioned on $i = i_0$  is at most  $(T+1)/2^{q+i_0} < (T+1)/2^q$ (which is necessary for her to win),
and then by the union bound, the probability of success for the
adversary in this case  will be at most $k(T+1)/2^q = O(T/\binom{k}{ck})$.
The reason is that the adversary can ask at most $T$ oracle queries after we reveal in order to find a
preimage for $\cO(x_{i_0})$ . The probability that for one of the queries $x$ among these $T$ queries  she asks we have
$\cO(x_i)=\cO(x)$ is at most $(T)/2^{q+i_0}$, and when it does not happen, the adversary has to guess a preimage for $\cO(x_i)$ blindly,
which will be correct with probability $1/2^{q+i_0}$.

\end{proof}

The constant $c$ in the description of the scheme maximizes $k
\choose{ck}$, conditioned on $q \approx  (1+c)k$. The same ideas
show that whenever $n \leq d q$ where $d \approx 0.812$ is
obtained as above ($d=H(c)/(1+c)$), then there is a one-time
signature scheme for messages $\bits^n$ that makes only $q$
queries and achieved security exponential in the length of its
queries.

\nnspace\Section{Extensions} \label{sec:extensions}

Now we prove several extensions of Theorem~\ref{thm:mainintro}.

\ifnum\full=0 \vspace{1ex}\noindent\textsc{Other oracles.} \else
\SubSection{Other oracles} \fi Using minor changes to the proof
of \theoremref{thm:main} we can get a similar lower bound for
signature schemes based on the ideal cipher or a random
permutation oracles. This is important as these oracles are also
sometimes used to model highly efficient symmetric-crypto
primitives, and so it is an interesting question whether such
oracles can be used to construct signatures more efficiently.

\begin{theorem}\label{thm:idealcipher}
Let $\cO$ be either the ideal cipher oracle. Then, for every one-time signature scheme for messages $\bits^n$ using
a total of $q \leq n/4$ queries to $\cO$ there is an adversary making
$2^{(4-o(1))q}$ oracle queries that breaks the scheme with
probability $1-o(1)$, where $o(1)$ denotes a term tending to $0$
with $q$. In case of  $\cO$ being the random permutation oracle, only $q \leq n/2$  is needed to get and adversary asking $2^{(2-o(1))q}$
queries, breaking the scheme with probability $1-o(1)$.
\end{theorem}

\ifnum\full=0 \nnspace\paragraph{Proof idea.} The proof for the
ideal cipher oracle is more challenging than random permutations,
and we explain how to handle this case. The crucial property of the
random oracle used in the proof of Theorem~\ref{thm:main} is that
the answer at every point is computed using independent randomness
from the answer at all other points. This is no longer true in the
ideal-cipher oracle since \textbf{(1)} if $y$ is the answer for the
query $(k,x,\mathsf{F})$ then $x$ must be the answer for
$(k,y,\mathsf{B})$ and \textbf{(2)} it cannot be the case that
$\cO(k,x,\mathsf{F}) = \cO(k,x',\mathsf{F})$ for some $x \neq x'$.
We handle \textbf{(1)} by stipulating that all algorithms, when
making a query of the form $(k,x,\mathsf{F})$, will make also the
redundant query $(k,y,\mathsf{B})$ and vice versa. This increases
the number of queries by at most a factor of $2$ but now allows us
to answer all (non-redundant) queries using independent randomness.
Issue \textbf{(2)} is not a problem if the space of possible $x$'s
is large enough (say at least $2^{3q}$) since then we can
approximate the random permutation by a random function. But if the
space is smaller than $2^{3q}$ then the adversary can simply query
all of these points, hence making the oracle meaningless.

\else

\begin{proof} We explain the proof for the ideal cipher oracle.
Extending the proof for the random permutation oracle is
straightforward.

We change both the signature scheme and the oracle for the sake of
the analysis. We let the new oracle $\cO'$ be the same as $\cO$
except that $\cO'$ does not answer queries of the form $(k,x,d)$
whenever $\abs{x} < 2(q+\log q)$. Instead it answers queries of the
type $(k,n)$ where $n < 2(q+ \log q)$, to which it returns the long
string containing the concatenation of $\cO(k,x,\mathsf{F})$ for $x
\in \bits^n$.

We change the signature scheme to get a new scheme $(\Gen', \Sign',
\Ver')$ as follows: \textbf{(1)} use $\cO'$ instead of $\cO$ and
\textbf{(2)} whenever an algorithm makes a query $(k,x,d)$ and
obtains an answer $y$, it will also make the ``redundant'' query
$(k,y,\bar{d})$ (where $\bar{d}=\mathsf{B}$ if $d=\mathsf{F}$ and
vice versa). %We call this redundant query the \emph{dual} of the original query.
Note that the total number of queries of the new
scheme is at most $q'=2q$.

\begin{lemma} \label{lem:changedWorld}
Given the scheme $(\Gen', \Sign', \Ver')$, there is an adversary
$\Adv$ making at most $\poly(q')2^{q'}$ queries from $\cO'$ that
breaks the scheme with probability $1-o(1)$.
\end{lemma}

Lemma~\ref{lem:changedWorld} implies Theorem~\ref{thm:idealcipher}
since any such adversary can be implemented using the oracle $\cO$
with at most a $q^2 2^{2q} $ factor increase in the number of
queries, and the total number of queries will be $\poly(q')2^{q'}
q^2 2^{2q} = 2^{(4-o(1))q}$.

\begin{proof}
The description of the attack remains basically the same as that of
\theoremref{thm:main} set by parameters in Corollary~\ref{cor:main} (i.e. $N = 2^q, \lambda = \delta = \theta(q^{-1/2})$), and we have the same distributions
$\Hyb^0,\Hyb^1,\Hyb^2,\Hyb^3$ as before. However, there are some
minor changes as follows:

\begin{itemize}
\item During Step~2 of the attack, whenever learn a query, we add both
the query and its dual to $L_i$.

\item During Step 4 of the attach we might discover an inconsistency between the
guesses we made in the sampled transcript $\Tilde{T}_0$ and the
answers we receive from the oracle $\cO$. That is, we might get the
same answer for two different plain texts with the same key.
However, as we will see this will only happen with small
probability, and we ignore this case safely.  .

\item The definition of $\Hyb^2$ needs to change a little.
Namely, in the experiment for the distribution $\Hyb^2$, during the
signing and verification of $\alpha_1, \dots, \alpha_N$, whenever we
make a new non-redundant query $(k,x,d)$, we look at all queries of
the form $(k, \cdot , d)$ appearing either in the transcript of the
system so far (i.e. $\Tilde{T_0}$, $T_1$, $\dots$) or in the learned
queries of $L_M$. Then we choose a random answer $y$ from the set of
unused answers and use it as the oracle answer for $(k,x,d)$. The
next redundant query $(k,y, \bar{d})$ is simply answered by $x$.

\end{itemize}

The differences between the proof in this case and the proof of
Theorem~\ref{thm:main} are the following:

\begin{itemize}

\item We need to include the condition in the event $E_j$
that the queries made in the $j^{\tth}$ signing and verification are
consistent with (the key generation part of) the transcript
$\Tilde{T}_0$ in the sense that they do not specify two queries
$(k,x,d), (k,x',d)$, $x \neq x'$ which map to the same answer $y$.
The consistency condition guarantees (by definition) that if $E_j$
occurs, then the verifier will accept the $j^{\tth}$ signature.

The combinatorial condition $V_j \cap (\Tilde{G} \cup \Tilde{S}_0)
\subseteq V_0$ still guarantees that the $j^{\tth}$ verification
does not ask any query for which we have guessed the
answer.\footnote{This also guarantees that there is no inconsistency
between the $j^{\tth}$ verification and the transcript
$\Tilde{T}_0$, but later we will show that the total consistency
happens with good probability}

We can still prove that   $\Pr_{\Hyb^0}[ \exists_j E_j ] =
\Pr_{\Hyb^1}[ \exists_j E_j ]$ using basically the same proof as in
\claimref{claim:T0T1}. We just have to note that as long as $E_j$
happens in both experiments, there is no way to distinguish their
$j\tth$ signing and verification, and the consistency also happens
either in both or in none of them.
s
\item We again show $\SD(\Hyb^1, \Hyb^2) = o(1)$. The reason is that the difference between the
    distributions $\Hyb^1$ and $\Hyb^2$ is due to some events
    which happen with probability $o(1)$. That is there are events in the experiments of sampling
    $\Hyb^1$ and $\Hyb^2$ which happen with probability $o(1)$ and conditioned on they not happening, $\Hyb^1$ and $\Hyb^2$ have the same
    distribution.

\begin{itemize}
\item Similar to \claimref{claim:T1T2} one of the differences between the distributions
$\Hyb^1, \Hyb^2$ might be because of $\Adv$ asking a query in
$Q(T_0)\setminus Q(L_M)$. Because of the same analysis given in the
proof of \claimref{claim:T1T2} the probability that we ask any such
query (in both experiments) is at most $2 \delta = o(1)$. So, in the
following we assume that this case does not happen.

\item In experiment of sampling $\Hyb^1$, when a new non-redundant query
$(k,x,d)$ is asked in the $1 \leq i\tth$ signing or verification,
the returned answer $y$ might be equal to a guessed answer for a
query $(k,x',d)$ of $\Tilde{T_0}$ (we call this event $F_1$), but it
is never equal to the answer of a query $(k,x'',d) \in Q(T_0)
\setminus Q(L_M)$. The situation for $\Hyb^2$ is the reverse: On a
new non-redundant query $(k,x,d)$ during the $1\leq i\tth$ signing
or verification, the answer is never equal to a guessed answer for a
query $(k,x',d)$ in $\Tilde{T_0}$, but it might be equal to the
answer of a query $(k,x'',d) \in Q(T_0) \setminus Q(L_M)$ (we call
this event $F_2$). Note that $(\Hyb^1 \mid \neg {F_1} )\equiv (\Hyb^2
\mid \neg {F_2} )$. As we will see, $\Pr[F_i] = o(1)$ for $i = 1, 2$
which shows that $\SD(\Hyb^1, \Hyb^2) = o(1)$.

The reason for $\Pr[F_1] = o(1)$ is that whenever we have a new
non-redundant query in the $1\leq i\tth$ signing or verification,
its answer is chosen from a set of size at least $q^2 2^{2q} -
q'2^{q'}$ which might hit a guessed answer for a query in
$\Tilde{T_0}$ with probability at most $q'/(q^2 2^{2q} - q'2^{q'}) =
o(1)$. The same argument holds for $\Pr[F_2] = o(1)$.

\end{itemize}

\item \claimref{claim:T2T3}  still holds with the similar
    proof because of the way we defined $\Hyb^2$ for the case of
    ideal cipher.

\item \claimref{claim:T3} is still correct with the same
    proof. Note that all the signing and verifications are
    consistent.

\end{itemize}

\end{proof}

A similar and simpler proof works for the case of a
random permutation oracle. In this case, we again change the oracle
by merging small queries into a single query with a huge answer, but
we don't have the issue of adding ``dual'' queries, and therefore
the condition $q \leq n/2$ (rather than
$q\leq n/4)$ is enough to get an adversary who breaks the scheme with probability $1-o(1)$ by asking $2^{(2-o(1))q}$
queries (rather than $2^{(4-o(1))q}$ queries).
\end{proof}
\fi

\ifnum\full=0 \vspace{1ex}\noindent\textsc{Implementing adversary in
$\BPP^{\NP}$} \else \SubSection{Implementing Adversary in
$\BPP^{\NP}$.} \fi If the signature scheme is efficient, using an
$\NP$ oracle, our adversary can run in time $\poly(n,2^q)$, where
$n$ is the length of messages to be signed.\footnote{In general, the
security parameter could be different from the length of the
messages $n$. For example, in Section~\ref{app:upperbound}, the
security parameter was $q$ (so the security was $2^{\Omega(q)}$),
and the running time of the algorithms was $\poly(n,q)$. Here, for
simplicity, we assume that $\ell = \poly(n)$, and all the
algorithms' queries are of length $\ell$.} That is, we prove the
following lemma:

\begin{lemma}\label{lem:bppnp}
If the signature scheme is efficient, the adversary of the proof of
Theorem~\ref{thm:main} can be implemented in $\poly(n,2^q)$ time
using an oracle to an $\NP$-complete problem.
\end{lemma}

Lemma~\ref{lem:bppnp} can be interpreted as saying that a non-black-box proof of security for a signature scheme more efficient than the
lower bounds provided by Theorem~\ref{thm:main} will necessarily imply a
proof that $\P \neq \NP$.

\ifnum\full=0 \nnspace\paragraph{Proof idea.}  $\Adv$ uses its
unbounded computational power to sample elements from certain
distributions, and find strings that are ``heavy'' with respect
to these distributions. It turns out these tasks can be
performed using the works on sampling witnesses with $\NP$
oracle \cite{JerrumVaVa86,BellareGoPe00}.

\else

\noindent The only place in which the adversary uses its unbounded
computational power is in Step~2 where it chooses $x_i$ to be the
lexicographically first unlearned string in $\bits^l$ such that
$x_i$ is queried in $\mathbf{D}_i$ with probability at least $\e$,
and in Step~3 when it samples a random $\Tilde{T}_0$ from
$\mathbf{D}_M$.

We show that:
\begin{itemize}
\item Using an $\NP$ oracle, we can sample from a
    distribution $\mathbf{D}'_i$ in  expected
$\poly(n,2^q)$ time such that $\SD(\mathbf{D}'_i, \mathbf{D}_i) \leq
\e$, where $\e$ is as defined in Step~2.

\item Using  the $\mathbf{D}'_i$  sampler, we
    can implement the adversary in  $\poly(n,2^q)$ time
    with similar success probability.

\end{itemize}

We first show how to use a $\mathbf{D}'_i$ sampler to implement the
adversary efficiently and then will show how to sample from
$\mathbf{D}'_i$ efficiently  using an $\NP$ oracle.

\paragraph{Efficient adversary using a  $\mathbf{D}_i$
approximate-sampler.} So, here we assume that we can sample
efficiently  from a distribution $\mathbf{D}'_i$ such that $
\SD(\mathbf{D}'_i, \mathbf{D}_i) \leq \e$. In order to choose $x_i$
in the $i\tth$ step of the learning phase, we do the following. Let
$m = (l + \log M - \log \delta)/\e^2$. We sample $m$ times from $\mathbf{D}'_i$ to get
$D_i^1, \dots, D_i^m$. Then we choose $x_i$ to be the
lexicographically first unlearned query (i.e. not in $L_{i-1}$)
which appears in at least $2\e$ fraction of $Q(D_i^j)$'s.

\begin{claim} \label{claim:goodEvents}
With probability at least $1-\delta$ we get the following:
For every $x \in \bits^l$, and every $1 \leq i \leq M$:

\begin{enumerate}
\item If $\Pr[x \in Q(\mathbf{D}_i)] \geq 3 \e$, then $x$ appears in
more than $2\e$ fraction of $Q(D_i^j)$'s.

\item If $\Pr[x \in Q(\mathbf{D}_i)] \leq  \e$, then $x$ appears in
less than $2\e$ fraction of $Q(D_i^j)$'s.

\end{enumerate}
\end{claim}

If the event above happens, it means that the learning algorithm
learns all the $3\e$-heavy queries in its $M$ rounds with
probability at least $1-\delta$ (using the same argument as before).
Therefore we get a weaker, yet strong enough, version of
\claimref{claim:T2T3} saying that the $\SD(\Hyb^1, \Hyb^2) \leq 3 \delta
+  \delta + \delta= o(\delta)$.

The Claim~\ref{claim:goodEvents} follows from the Chernoff bound.
The probability that any specific $x$ violates the claim's condition
in any of the rounds is at most $e^{-m\e^2}<2^{-m\e^2} = 2^{-l - \log M + \log \delta}$.
By union bound, the probability that the event is not violated at most $M 2^l 2^{-l - \log M + \log \delta} = \delta$.

\paragraph{Sampling  $\mathbf{D}'_i$ efficiently using an
$\NP$ oracle.}

Note that $L_i$ which captures our knowledge of the system after the
$i\tth$ round of the learning phase can be encoded with
$\poly(n,2^q)$ bits. The number of random bits used by the adversary
till the end of the $i\tth$ round of the learning phase is also
$\poly(n,2^q)$. For some technical reason which will be clear later,
we add the randomness used by the adversary to the description of
$L_i$. Similarly, any (possible) transcript $D$ which
$\Pr[\mathbf{D}_i = D]
>0$  can be represented with $\poly(n,q) < \poly(n,2^q)$ bits. In
the following we always assume that such encodings are used to
represent $L_i$ and $D$.

In order to sample from a distribution close to $\mathbf{D}_i$ we
use the following Lemma:

\begin{lemma} \label{lem:counting}
There is a function $f: \bits^* \times \bits^* \mapsto \N$
which is efficiently computable (i.e. time $\poly(n, 2^q)$), with the following
properties:

\begin{enumerate}
\item $f(L_i, D)  =  \floor{  c\P[ \mathbf{D_i} = D ]} $ for some
    constant $c$  depending on $L_i$. So we have $f(L_i, D) = 0$ if $\Pr[\mathbf{D_i} = D] = 0$.
    \Mnote{ writing it in form of $\in c\Pr[.] \pm 1$ is wrong, because then for zero probabilities
    we might output $f=1$ and there are a lot of them.}

\item $f(L_i, D) \geq 10/\e$ whenever $\Pr[\mathbf{D} = D] > 0$
where $\e$ is as defined in Step~2.
\end{enumerate}

\end{lemma}

\Mnote{We have to feed $L_i$ to f, because we want its running time
to be a fixed $\poly(n,2^q)$ independent of $i$}.

Before proving the lemma, we see how it is used.

\begin{corollary}
We can sample from a distribution $\mathbf{D}'_i$ such that
$\SD(\mathbf{D}_i, \mathbf{D}'_i) \leq \e$ in time $\poly(n,2^q)$ (where the time $\poly(n, 2^q)$ is independent of $i$ for $1 \leq i \leq M$).
\end{corollary}

\begin{proof}
Let $W_i = \{(D, j) \mid 1 \leq j \leq f(L_i, D) \}$ be the set of
``witnesses'' for $L_i$, where $f$ is the function in
Lemma~\ref{lem:counting}. Lemma~\ref{lem:counting} shows that the
relation $R = \{(L_i, w) \mid w \in W_i \}$ is an $\NP$ relation. It
is known \cite{BellareGoPe00} that for any $\NP$ relation, there is
a witness-sampling algorithm that given any $x$, samples one of the
witnesses of $x$ uniformly in expected $\poly(\abs{x})$ time.
Therefore, we sample a random $w = (D,j)$ such that $w \in W_i$ in
expected $\poly(n,2^q)$-time, and output $D$.  It is easy to that the
distribution $\mathbf{D'_i}$ of our output has statistical
distance at most $\e$ from the distribution $\mathbf{D_i}$.
\end{proof}

\begin{proof}(Lemma~\ref{lem:counting})
Recall that $\mathbf{D}_i$ is the distribution of transcripts $T_0$
conditioned on the information given in $L_i$. Let the event
$E(L_i)$ be the event that during the running of the system (and our
attack) adversary's knowledge about the system and its randomness
after the $i\tth$ round of the learning is what $L_i$ denotes.
Similarly, let $E(D)$ be the event that $D=T_0$ is the case in our experiment. Thus, for every transcript $D$, $\Pr[ \mathbf{D}_i = D] =
\Pr[ E(D) \mid E(L_i) ]$. If we could compute $\Pr[ E(D) \mid E(L_i) ]$,
we could somehow use it in the Lemma~\ref{lem:counting}, but instead
of doing that, we will rather compute $\Pr[E(D) \wedge E(L_i)]$ which is
proportional to $\Pr[ E(D) | E(L_i) ]$ up to a constant factor depending on
$L_i$, and will scale it up to some big integer.

Given $L_i$ and $D$, in order to compute $\Pr[E(D) \wedge E(L_i)]$,
we track the whole experiment from the beginning in the following order:

\begin{itemize}
\item Key Generation
\item Signing $\alpha_0$
\item The attack (which includes the verification of $\alpha_0$ as
its first step) to the end of the $i\tth$ round of the Learning.
\end{itemize}

At any moment that some  coin tossing is involved  (i.e. in the key
generation algorithm, in the attack, or fin an oracle answer), the result is determined by the description
of $L_i$ and $D$. Thus, we can calculate the probability that given values
of $L_i$ and $D$ will be the ones in the real
running of the experiment\footnote{For the case of ideal cipher or random permutation oracles, we need to keep track
of the oracle answers so far during the simulation of the experiment, in order to know what is that probability of receiving
a specific answer from the oracle at any point.}. More quantitatively, during the simulation of the experiment,
we  receive any specific oracle answer with probability at least $2^{-l}$ whenever
it is a possible answer and the probability of getting a specific random tape for the key generation and the adversary is
at least  $2^{-\poly(n,2^q)}$. Since the total probability of
$\Pr[E(D) \wedge E(L_i)]$ is the multiplication of all those
probabilities that we get during the simulation of the system, and because
the number of oracle queries that we examine is at most $2^{O(q)}$, we get
$\Pr[E(D) \wedge E(L_i)] > 2^{-\poly(n,2^q)}$ whenever $\Pr[E(D)
\wedge E(L_i)] \neq 0$. Note that $\e$ in the attack is $2^{-O(q)}$.
Therefore, for a big enough constant $c = \poly(n, 2^q)$, the
function $f(L_i,D) = \floor{c \Pr[E(D) \wedge E(L_i)]}$ is
computable in time $\poly(n,2^q)$ and  we have $f(L_i,D) > 10 / \e$ as well.
\end{proof}

\fi

\ifnum\full=0 \vspace{1ex}\noindent\textsc{Handling imperfect
completeness} \else \SubSection{Handling imperfect completeness}
\fi \label{subsec:ImpComp} While the typical definition of a signature scheme
stipulates that a valid signature (generated by the signing
algorithm with the correct key) should be accepted with
probability one, it makes sense to consider (especially in the
context of negative results) also signatures where the verifier
may reject such signatures with small probability, say $1/10$.
We are able to extend our result to this case as well:

\begin{theorem}\label{thm:imperfect}
For every one-time signature scheme for messages $\bits^n$, accepting correct
signatures with probability at least $0.9$ (over the randomness of the verifier),
and asking a total  $q \leq \sqrt{n}/20$ queries to a random oracle, there
is \textbf{(1)} an adversary making $2^{(1+o(1))q}$ queries that
breaks the scheme with probability at least $2^{-q}$ and \textbf{(2)} an
adversary making $2^{O(q^2)}$ oracle queries that breaks the scheme
with constant probability.
\end{theorem}

\ifnum\full=0 \nnspace\paragraph{Proof Idea.} The proof of
Theorem~\ref{thm:main} easily extends to show \textbf{(1)}: one
can show that the adversary $\Adv$ of that proof will obtain a
collection $2^q$ signatures such that for every fixing of the
verifier's random tape, one of these signatures will be
accepted, implying that if we choose a signature at random it
will be accepted with probability $2^{-q}$. However, this is
somewhat unsatisfying for a lower bound, since arguably an
attack that succeeds with very small probability is much less
devastating than an attack that succeeds with probability close
to $1$. To prove \textbf{(2)}  we need to use a more careful
argument (and lose a quadratic factor), applying the
combinatorial lemma to the sets of queries that are ``heavy''
(asked with probability at least $1/(10q)$) by the verifier, and
using a smaller value of $\e$ in Step~2 of the attack.

\else

The proof of part \textbf{(1)} is a straightforward extension of the
proof of Theorem~\ref{thm:main} and so we bring here the proof of
part \textbf{(2)}:

\begin{lemma}
For every one-time signature scheme with imperfect completeness
(i.e., verifier can reject valid signatures with probability at most
$1/10$ over its coins) there is an adversary asking $2^{O(q^2)}$
queries that finds with probability $1-o(1)$ a message/signature
pair which passes the verification with probability at least $0.7$. \footnote{ The probability $0.7$ could be substituted by any constant less than
0.9 with changing the constants in the proof.}
\end{lemma}

\begin{proof} The main difference between the proof of this lemma compared to that of
Theorem~\ref{thm:main} is the way we define the sets $V_j$'s. They are not simply the queries that the verifications ask from the oracle.
For sake of analysis, for every $j$, we define the set $V_j$ to be
the set computed by the following process: run the $j\tth$
verification algorithm on the generated message/signature pair
$m=$ times (for $m$ to be defined later), and let $V_j$ be the set of queries that appeared
in at least a $1/(20q)$ fraction of these verifications. Hence, we
have $|V_j| \leq 20q^2$. Note that the definition of $V_j$ depends
on the oracle used to do the verifications.  We will treat the sets $V_j$'s in the analysis similar to what we did to them with their previous definition.
So, we define the new parameter $r = 20q^2$ to the upper bound on $|G| + |S_j| + |V_j|$, while $q$ is still an upper bound for $|G| + |S_j|$.
As we will see, the proof will be similar to that of Theorem~\ref{thm:main} and the parameters are set similar
to those of Corollary~\ref{cor:main}: $N = 2^{r}, \lambda = \delta = \frac{{r \choose r/2}}{2^{r}} = \theta(r^{-1/2}) = \theta(1/q), m = 20^3 q^4,
\e = \frac{\delta}{mqN}
,M = \frac{q}{\e \delta}$.
Other than the parameters, the differences compared to the previous attack are:

\begin{enumerate}

\item When obtaining the signature $\sigma_0$ in Step~1, we
    run the verification algorithm $m$ times and
    record in $L_0$ all the resulting query/answer pairs.

\item In Step~4 we test $q^3$
    times each generated message/signature pair and output the first signature that
    passes the verification at least a $0.75$ fraction of
    these $q^3$ times.

\end{enumerate}

We also define the set
$U_j$ to be the set of queries that the $j\tth$ verification asks
from the oracle with probability at least $1/(10q)$ over its own
randomness after we fix the random oracle. Hence we have  $|U_j|
\leq 10q^2$

We say that $E_j$ holds if (as before) $V_j \subseteq (\Tilde{G}
\cup \Tilde{S}_0) \cap V_0$. We also say that the event $E$ holds if $U_j \subset V_j$ for every $j$.

\begin{claim} If $E_j \wedge E$ holds, then the $j^{\tth}$ signature will be accepted
by the verifier with probability at least $0.9-0.1=0.8$ over the
randomness of the verifier.
\end{claim}

\begin{proof} \label{claim:notMissing}
The only way this won't happen is that with probability at least
$1/10$, the verifier makes a query in the (at most $q$-sized) set
$\Tilde{G} \cup \Tilde{S}_0 \setminus V_0$. But if this happens,
then there is a query in that set that is queried first with
probability at least $1/(10q)$, yet because $E$ holds that means
that it will be contained in $U_j \subset V_j$, contradicting $E_j$.
\end{proof}

For any specific $1 \leq j < N$, by Chernoff bound, the probability that the fraction of times that we accept the generated signature for
$\alpha_j$ is $0.05$ far from its real probability of being accepted by the verifier is at most
$e^{-0.05^2 q^3}$ and by union bound, the probability that it happens for some $j$  is at most $  2^{20q^2} e^{-q^3/400} = o(1)$.
Now suppose $E_j \wedge E$ holds for some $j = j_0$. So by Claim~\ref{claim:notMissing}, fistly we will output a pair of message and signature,
and secondly this pair is accepted by the verifier with probability at least $0.7$.

\begin{claim}
We have $\Pr[E] \geq 1-o(1)$.
\end{claim}

\begin{proof}
By the Chernoff bound, the probability that a particular member of
$U_j$ is not in $V_j$ is at most $e^{-(\frac{1}{20q})^2m} =
e^{-20q^2}$. By union bound over the members of $U_j$, and $j$ we
have $\Pr[\textbf{(2)}\text{ fails for some }j] \leq 10q^2 2^{20q^2}
e^{-20q^2} = o(1)$.
\end{proof}

Now that we know $E$ holds almost always, it only remains to show that with high probability $E_j$ happens
for some $j$. This time we define the four hybrid distributions
$\Hyb^0,\Hyb^1,\Hyb^2,\Hyb^3$ a bit different. Instead of putting in
$\Hyb^i$ the query/answer pairs that we received during  one
verification, we put in $\Hyb^i$ all such pairs that we get at some
point during the $m$ times that we run the verification.

The proofs of Claims~\ref{claim:T0T1}--\ref{claim:T3} also work
basically in the same way as before:

\begin{itemize}
\item \claimref{claim:T0T1} still holds with the same proof.

\item \claimref{claim:T1T2} still holds with the same proof because of the new
smaller value of $\e$ that we used.

\item \claimref{claim:T2T3}  still holds with the same proof.

\item \claimref{claim:T3} is still correct with the same
    proof because the condition $q \leq \sqrt{n} /20$  guarantees that there is enough
    room to choose $N \leq 2^n$ different messages in the attack.

\end{itemize}

So, our adversary asks at most $Nmq + M + Nq^3 = \poly(q)2^r = 2^{O(q^2)}$ queries, and with probability $1-o(1)$ finds a pair of message/signature
passing the verification with probability at least $0.7$.
\end{proof}

\fi

We note that the combination of all the above extensions holds
as well (e.g., we can implement in $\BPP^{\NP}$ an adversary
that breaks any signature scheme with imperfect completeness
that is based on the ideal cipher).

\ifnum\full=0 \vspace{1ex}\noindent\textsc{Efficiency of the
verifier} \else \SubSection{Efficiency of the verifier} \fi Because
the signing and verification algorithms are run more often than the
key generation, lower bounds on their own efficiency is still
meaningful. In Section~\ref{app:upperbound} we saw that the signing
algorithm can be very efficient while the total number of queries
was almost optimal. Here we show that if we want to get an efficient
verifier and exponential security at the same time, it makes the
total number of queries to be inefficient.

\begin{theorem}\label{thm:tradeoff} For every one-time signature
scheme for messages $\mathcal{ M}$ with total $q$ oracle queries where, if the verification asks at
most $v$, $v \leq q/2$ oracle queries and $|\mathcal{M}| \geq \frac{{q \choose v}}{\lambda}$ then there
is an adversary asking at most $O(\frac{q^2 {q \choose v} }{\lambda \delta^2})$ queries that
breaks the scheme with probability at least $1-\lambda-\delta$.
\end{theorem}

Before going over the proof note that for any $v,k\in \N$,  where $3 \leq v \leq \frac{q}{2}$ (i.e. $1 \leq v-2 \leq k$ where $v+k+2=q$) the scheme of
Section~\ref{app:upperbound} can be simply changed to get a new scheme
in which the verifier asks  $v$ queries by revealing $v-2$ sized
subsets of $x_i$'s as the signature rather than $ck$ sized ones. A similar proof to that of Theorem~\ref{thm:LamChanged} shows that this
new scheme has security $\Omega(\binom{k}{v-2}) = \Omega(\binom{q-v-2}{v-2})$. So, if $v = dq$ for constant $d$, the maximum security $S$ one
can get by asking at most $v= q/d$ queries in verification and $q$ queries totally is bounded as
$H(\frac{1}{d-1})(1-1/d)-o(1) \leq \frac{\log S}{q} \leq H(\frac{1}{d}) + o(1)$ where $H(\cdot)$ is the Shannon's entropy function and $o(1)$ goes
to zero with $q$.

\begin{proof}(Theorem~\ref{thm:tradeoff}) The proof is almost the same as that of
Theorem~\ref{thm:main}. The only difference is in Claim~\ref{claim:T3} in which we have a restriction that $|V_j| \leq v$, and we conclude that
$K \leq {q \choose v}$. The only difference in the proof of Claim~\ref{claim:T3} is that now the event $A_i$ has probability at least
$\tfrac{|U_i|!|V_i|!}{(|U_i|+|V_i|)!} =
1/\binom{|U_i|+|V_i|}{|V_i|} \geq 1/\binom{q}{|V_i|} \geq
1/\binom{q}{v}$ because $v \leq q/2$.
\end{proof}

\Section{Lower bounds on black-box constructions}
\label{subsec:BBconst}

In a construction for signature schemes, one might use a
\emph{standard primitive} (e.g., one way function) rather than one with ideal security (e.g., random function).
These constructions could have different levels of
``black-boxness'' discussed thoroughly in \cite{ReingoldTrVa04}. What we will call black-box, is called fully
black-box in \cite{ReingoldTrVa04}. Here we give a more quantitative definition  of such constructions. For simplicity we only define
the black-box constructions of signature schemes from hard one-way functions, and the others are similar.
After giving the formal definitions we will prove strong lower bounds on the
efficiency of signature schemes from symmetric primitives when the construction is black-box.

\begin{definition} \label{def:harfFunc}
Let $F_{\ell}$ denote the set of all functions $f \colon \bits^{\ell} \to \bits^{\ell}$ over $\ell$ bits.
 We call a family of functions $\{f_{\ell} \mid \ell \in \N, f_{\ell} \in F_{\ell} \}$, \emph{$s$-hard} (to invert), if for any
 probabilistic algorithm $A$ running
in time at most $s(\ell)$, we have $\Pr_{x \gets_R \bits^{\ell}}[A(f(x)) \in f^{-1}(f(x))] \leq \frac{1}{s(\ell)}$ where the probability is over
the choice of $x$ and the coin tosses of $A$.
\end{definition}

By $S$-hard functions, for a \emph{set} of functions $S$, we mean all those which  are $s$-hard for some $s \in S$.
(Think
of $S$ as the set of all the functions which are super-polynomial, quasi-polynomial, or exponential etc...) So, we will
keep the notation that the capital
$S$ denotes a set of functions.

For simplicity we use $n$, the length of the messages to be signed, as the security parameter of the signature scheme (i.e, the efficient schemes
will run in time $\poly(n)$ and for larger values of $n$ the scheme becomes more secure).

\nnspace\begin{definition} \label{def:BBsig}
A  \emph{black-box}
construction of one-time signature schemes for $n$-bit messages  from $S$-hard one-way functions, with security parameter contraction $\ell (n)$
is made of the following two families of reductions for all $n \in \N$:

\begin{itemize}
\item The implementation reduction  $I = (\Gen, \Sign, \Ver)$ has three components which are algorithms running in time $ \poly(n)$ ($\Gen$ is
probabilistic) and $I^f = (\Gen^f, \Sign^f, \Ver^f)$ satisfies
in Definition~\ref{def:sig} by setting $\cO = f$ for any $f \in F_{\ell(n)}$.

\item We call $A$ a  \emph{$I^f$-breaker} if $A$ is a (not necessarily efficient) adversary who breaks the security of
$I^f$ with non-negligible probability over its own randomness
by playing in the game defined in Definition~\ref{def:sigsec}. The security reduction $R$ is an algorithm running in time $t(n)$ where:
\textbf{(1):}  For any $f\in F_{\ell(n)}$ and any $I^f$-breaker $A$,
$\Pr_{x \getsr \bits^{\ell(n)}} [R^{A, f}(f(x)) \in f^{-1} (f(x)) ] \geq \frac{1}{w(n)} $ where the probability is over
the choice of $x$ and the coin tosses of $R$ and $A$, \textbf{(2):} $t(n) p(n) < s(\ell(n))$ for any $p(n) = \poly(n)$, any  $s(\cdot) \in S$ and and large enough $n$, and \textbf{(3):}  $w(n) < s(\ell(n))$ for any $s\in S$ and large enough $n$.

\end{itemize}
\end{definition}

The security parameter contraction factor $\ell(n)$ in Definition~\ref{def:BBsig} measures how small the length of the function  used
in the reduction is (i.e., the security parameter of the primitive used)  compared to  $n$ (i.e., the security parameter of the signature scheme). The term ``security parameter expansion'' is used in \cite{HaitnerHoReSe07} for the inverse
of the contraction parameter.

Note that having such a reduction, the existence of any efficiently computable family of functions $f \colon \bits^{\ell} \to \bits^{\ell}$
which is $s$-hard to invert for some $s\in S$ implies the existence of (efficient) one-time signature schemes which are secure against polynomial-time adversaries.
That is because \textbf{(1):} We get an efficient implementation of the scheme by efficiently implementing $f$ for $I^f$, and \textbf{(2):}
If $A$ is a  $I^f$-breaker running in time $\poly(n)$, the reduction $R$ combined with its subroutine $A$ breaks the $s$-hardness of $f$ which
is not possible.

Now we prove a strong lower bound on the efficiency of signature schemes relying on the efficiency of strong one-way functions. Then we will show how it generalizes to any symmetric primitive and also functions with many hard-core bits.

\begin{theorem} \label{thm:lbFromOWF} Let $E$ denote the set of functions $E = \{f(\ell) \mid f = 2^{\Omega(\ell)} \}$.
Any black-box construction of one-time signature schemes for $n$-big messages from $E$-hard one-way functions with security parameter
contraction $\ell(n)$ needs to ask $\min(\Omega(\ell(n)), n)$ queries from the one-way function.
\end{theorem}

Before going over the proof we make two observations. First, if construction uses $E$-hard functions, it means that we should have $t(n) = 2^{o(\ell(n))}$ and $w(n) = 2^{-o(\ell(n)}$ in the security reduction. Another point is that the existence of such a reduction regardless of how many queries it asks, makes  $\ell(n)$ to be $\omega(\log n)$ for otherwise the condition $t(n) \poly(n) < s(\ell(n))$ in Definition~\ref{def:BBsig} will be violated. Therefore without loss of generality, we assume that $q \geq \log n$, because otherwise we can ask $\log n$ redundant queries in the key generation algorithm without changing the condition $q \leq \min(\Omega(\ell(n)), n)$.

\begin{proof} For sake of contradiction suppose that there is a black-box construction  of signature schemes $(I,R)$ where $I$ asks  $q \leq n$ queries from the one-way function and $\log n \leq q = o(\ell(n))$.

The proof will go in two steps. We will first show that any such construction results in a (computationally unbounded) adversary asking $2^{o(\ell)}$ queries from a
a random function $f \gets_R F_{\ell}$ and inverting it on a random point with probability at least $2^{-o(\ell)}$ (where this probability is also
over the choice of $f$). Then we will show that it is not possible to have such an adversary, namely
any adversary asking $2^{\ell/3}$ queries has chance of at most $2^{-\ell/3}$ for doing so.

\vspace{1ex} \noindent \textsc{Step 1.} Let $A$ be the adversary of Corollary~\ref{cor:main} for the implementation of the signature scheme $I$ (note $q \leq n$) asking at most $2^{(1+o(1))o(\ell(n))}$ queries from the function $f$ (note $q \leq o(\ell(n))$ and breaking $I^f$ with probability at least $1-o(1)$ when $f$ is chosen at random  $f \gets_R F_{\ell(n)}$ where $o(1)$ goes to zero with $q$. For large enough $\ell(n)$, $n$ becomes
large enough too, and so does $q$  (because $q \geq \log n$). Therefore $A$ asks at most $2^{o(\ell(n))}$ queries from $f$ and breaks the scheme with probability at least $3/4$ when $f \gets_R F_{\ell(n)}$ for large enough $\ell(n)$. By an average argument, with probability at least $1/2$ over the choice of $f$, $A$ breaks $I^f$ with probability at least $1/2$ over its own randomness. We call such $f$'s the good  ones. Whenever $f$ is good, $R^{A^f,f}$ inverts $f$ on a random point with probability at least $2^{-o(\ell(n))}$, and because $f$ is good with probability at least $1/2$, $R^{A,f}$ inverts $f$ on a random point with probability at least $2^{-o(\ell(n))}$ for a randomly chosen  $f \gets_R F_{\ell(n)}$ where the probability is over the choice of $f$, the choice of the image to be inverted, and the randomness of $A$. By merging the code of $R$ with $A$, we get an adversary $B = R^A$ who asks at most $2^{o(\ell(n))} 2^{o(\ell(n))} = 2^{o(\ell(n))}$ queries from $f \gets_R F_{\ell(n)}$ and inverts it on a random point (i.e., $y = f(x)$ for $x \gets_R \bits^{\ell(n)}$) with probability at least $2^{-o(\ell (n))}$.

\vspace{1ex} \noindent \textsc{Step 2.} Suppose $B$ is an adversary asking $2^{\ell/3}$ queries from a random  function $f \gets_R F_{\ell}$ trying to find a preimage for $f(x)$ where $x \gets_R \bits^{\ell}$. We can pretend that the value of $f$ at each point is determined at random whenever it is asked for the first time. So, at first $x$ is chosen, $f(x)$ is chosen, and it is given to $B$. At first $B$ does not have any information about $x$, so the probability that $B$ asks $x$ in any of its $2^{\ell/3}$ queries is at most $2^{-2 \ell /3}$. Assuming it does not ask $x$, the probability that $B$ receives the answer $f(y) = f(x)$ by asking any $y \neq x$ is at most $2^{-2 \ell /3}$. Assuming that none of the mentioned events happens, if it outputs $y$ differen from all queries it has asked from $f$, $f(y) = f(x)$ happens with probability $2^{-\ell}$. So its chance of winning is at most $2^{-2 \ell /3} + 2^{-2 \ell /3} + 2^{- \ell} < 2^{-\ell/3}$ (for $\ell \geq 4)$.

\end{proof}

As it is clear from  the theorem, our lower bound becomes stronger for larger values of $\ell(n)$ which is also the case in the similar (unconditional) lower bound results  \cite{HaitnerHoReSe07,Wee07}.

In order to extend the lower bound to other symmetric primitives (and functions with many hard-core bits) we can follow the same steps of the proof of
Theorem~\ref{thm:lbFromOWF} using the following lemma.

\begin{lemma} \label{lem:usingOracles}
 Let $P$ be a symmetric primitive (i.e, one-way function, one-way permutation, collision resistent hash function,  pesudorandom generator, pseodorandom function, message authentication code, or block cipher) , or the primitive  of functions $f \colon \bits^{\ell} \to \bits^{\ell}$ with $\ell/2$ hard-core bits. Then, there is an implementation for $P$ for security parameter $\ell$ with access to either, random oracle, random permutation oracle, or ideal cipher oracle which asks only a constant number of queries of length $\theta(\ell)$  from the oracle, and any (computationally unbounded) adversary $\Adv$ who asks at most $2^{o(\ell)}$ queries
 from the oracle has chance of at most $2^{-\Omega(\ell)}$ of breaking it (over the randomness of $\Adv$ and the oracle used).
\end{lemma}

\begin{proof} We will describe the natural implementations  and will show the proof of security only for the case that $P$ is the primitive of functions with $\ell/2$ hard-core bits. The security proofs for other implementations are also easy to get (in fact, we  already gave the proof for the case of one-way function in the proof of Theorem~\ref{thm:lbFromOWF}).

\begin{itemize}
\item \textbf{One-way function} using random oracle: To define the value of the function $f$ on input $x \in \bits^{\ell}$, we simply use the oracle's answer: $f(x) = \cO(x)$.

\item \textbf{One-way permutation} using random permutation oracle: To define  the value of the permutation $p$ on input $x \in \bits^{\ell}$, we simply use the oracle's answer: $p(x) = \cO(x)$.

\item \textbf{Collision resistent hash function} using random oracle: The value of the hash function $h$ on input $x \in \bits^{\ell}$ is made by using the first $\ell/2$ bits of the oracle's answer: $h(x) = b_1\dots b_{\ell/2}$ where $\cO(x) = b_1 \dots b_{\ell}$.

\item \textbf{Pseudorandom generator} using random oracle: The stretched output of the generator $g$ on input $x \in \bits^{\ell}$ is the output
of the oracle on the padded query: $g(x) = \cO(x|0^{\ell})$.

\item \textbf{Pseudorandom function} using random oracle: Using the key $k \in \bits^{\ell}$ on input $x \in \bits^{\ell}$, the output of
the function will be the first $\ell$ bits of the oracle's answer on the query made by attaching $k$ and $x$:  $f_k(x) = b_1 \dots b_{\ell}$ where $\cO(k|x) = b_1 \dots b_{2\ell}$.

\item \textbf{Message Authentication Code} using random oracle: Using the key $k \in \bits^{\ell}$, the authentication code of the message $x \in \bits^{\ell}$ is defined similar to that of pseudorandom function. The verification is clear.

\item \textbf{Block cipher} using ideal cipher oracle: Using the key $k \in \bits^{\ell}$ and the input $x \in \bits^{\ell}$, and the direction $d$ we simply use the oracle's answer $\cO(k,x,d)$ as our cipher.

\item \textbf{Function with $\ell/2$ hard-core bits} using random oracle: The value of the function $f$ on input $x = x_1 \dots x_{\ell}$ uses the oracle's answer: $f(x) = \cO(x)$ and the hard-core bits for $x$ will be the first $\ell/2$ bits of it: $HC(x) = x_1 \dots x_{\ell/2}$.

\end{itemize}

Now we prove the claim for the last primitive (i.e., functions with $\ell/2$ hard-core bits). Suppose the adversary $A$ asks at most  $2^{\ell/4}$ queries from the function $f$. Again, we assume that $f$ chooses its answers randomly whenever asked for the first time.  In order to break the hard-core property of the function $f$, the adversary $A$ needs to distinguish between two experiments. In the first one she is given $(f(x), U_{\ell/2})$ as in put, and in the second one she is  given $(f(x), HC(x))$, and in both of the experiments  $f \gets_R F_{\ell}$ and $x \gets_R \bits^{\ell}$ are chosen at random. Note that as long as $A$ does not ask $x$ from the oracle, the two experiments are the same. At the beginning $A$ does not knows the second half of the bits of $x$. So the probability that she asks $x$ from the oracle in one of her $2^{\ell/4}$ queries is at most $2^{\ell/4} 2^{-\ell/2} = 2^{-\ell/4}$. Hence, if the probability that she outputs $1$ in the experiment $i$ is $p_i$ (for $1 \leq i \leq 2$), we have $\abs{p_1 - p_2} \leq 2^{-\ell/4}$.

\end{proof}

So by using Lemma~\ref{lem:usingOracles} and following the steps of the proof of Theorem~\ref{thm:lbFromOWF} we get the following theorem:

\begin{theorem} \label{thm:lbFromOWF} Let $E$ denote the set of functions $E = \{f(\ell) \mid f = 2^{\Omega(\ell)} \}$, and $P$ be either
a symmetric primitive or the primitive of functions with $\ell/2$ hard-core bits.
Any black-box construction of one-time signature schemes for $n$-bit messages from an $E$-hard primitive $P$ with security parameter
contraction $\ell(n)$ needs to ask $\min(\Omega(\ell(n)), n/4)$ queries from the primitive $P$.
\end{theorem}

\nnspace\Section{Conclusions and open questions}

We believe that lower bounds of this form--- the efficiency of
constructing various schemes using black box idealized primitives---
can give us important information on the efficiency and optimality
of various constructions. In particular, three  natural questions
related to this work are:

\begin{itemize}

\nnspace\item Can one pinpoint more precisely the optimal number of
queries in the construction of one-time signature schemes based on
random oracles? In particular, perhaps our lower bound can be
improved to show that the variant of Lamport's scheme given in
Section~\ref{app:upperbound} is optimal up to lower order terms.

\Mnote{I have added the next question. Do you want to leave the
previous question in this form? Maybe speaking directly about the
constant is better.}

\nnspace\item What is the threshold $d$ that whenever $n \leq dq$,
we can get signature schemes for messages $\bits^n$ using $q$ oracle
queries, and arbitrary large security? Again, it seems that the
variant of Lamport's scheme given in Section~\ref{app:upperbound} (working for $ \log {k \choose ck}$ bit messages without hashing)
gives this threshold (i.e., $d \approx 0.812$).

\nnspace\item Can we obtain a $2^{O(q)}$-query  attack succeeding
with high probability against signature schemes with imperfect
completeness?

\nnspace\item Are there stronger bounds for \emph{general} (not
one-time) signatures? A plausible conjecture is that obtaining a
$T$-time signature with  black-box security $ S$ requires
$\Omega(\log T \log S)$ queries.

\end{itemize}

\vspace{1ex}\noindent\textbf{Acknowledgements:} We thank David
Xiao for useful discussions.

\nocite{GennaroTr00,GennaroGeKa03}

 \ifnum\full=1
\bibliographystyle{alphasy}
\else
\bibliographystyle{latex8}
\fi
%%%%%%%%%%%%%%%%%%%%%%%%%%%%%%%%%%%%%%%%%%%%%%%%%%%%%%%%%%%%%%%%%
% short alpha bib
% define the following commands in LaTeX file to change behavior:
% \biblkeystyle{..} :  style of bib key
% \biblbegin        :  stuff before the reference list
% \biblend          :  stuff before the reference list
%%%%%%%%%%%%%%%%%%%%%%%%%%%%%%%%%%%%%%%%%%%%%%%%%%%%%%%%%%%%%%%%%
\makeatletter
%%%%%%%%%%%%%%%%%%%%%%%%%%%%%%%%%%%%%%%%%%%%%%%%%%%%%%%%%%%%%
% following 3 lines might be problematic if not using Latex2e
\providecommand{\biblkeystyle}[1]{#1}
\providecommand{\biblbegin}{}
\providecommand{\biblend}{}
\@ifundefined{beginbibabs}{\def\beginbibabs{\begin{quotation}\noindent}
\def\endbibabs{\end{quotation}}}{}
\makeatother

\end{document}